\documentclass[review]{elsarticle}
\def\*#1{\boldsymbol{#1}}
\usepackage{hyperref,amsmath,amssymb,graphicx,cases}
\journal{Submitted to Journal}

\begin{document}

\begin{frontmatter}

\title{Reconstruction Error Bounds for Compressed Sensing under Poisson Noise using the Square Root of the Jensen-Shannon Divergence}

%\author{Sukanya Patil and Ajit Rajwade\fnref{myfootnote}}

\author[mymainaddress]{Sukanya Patil}
\ead{sukanyapatil1993@gmail.com}
\address[mymainaddress]{Department of Electrical Engineering, IIT Bombay}

\author[mymainaddress1]{Karthik S. Gurumoorthy}
\ead{karthik.gurumoorthy@icst.res.in}
\address[mymainaddress1]{International Center for Theoretical Sciences, TIFR Bangalore}
\fntext[myfootnote]{Karthik S. Gurumoorthy thanks the AIRBUS Group Corporate Foundation Chair in Mathematics of Complex Systems established in ICTS-TIFR.}

\author[mymainaddress2]{Ajit Rajwade\corref{mycorrespondingauthor}}
\cortext[mycorrespondingauthor]{Corresponding author}
\ead{ajitvr@cse.iitb.ac.in}
\address[mymainaddress2]{Department of Computer Science and Engineering, IIT Bombay}
\fntext[myfootnote]{Ajit Rajwade gratefully acknowledges support from IIT Bombay Seed Grant number 14IRCCSG012.}

\begin{abstract}
Reconstruction error bounds in compressed sensing under Gaussian or uniform bounded noise do not translate easily to the case of Poisson noise. Reasons for this include the signal dependent nature of Poisson noise, and also the fact that the negative log likelihood in case of a Poisson distribution (which is directly related to the generalized Kullback-Leibler divergence) is not a metric and does not obey the triangle inequality. There exist prior theoretical results in the form of provable error bounds for computationally tractable estimators for compressed sensing problems under Poisson noise. However, these results do not apply to \emph{realistic} compressive systems, which must obey some crucial constraints such as non-negativity and flux preservation. On the other hand, there exist provable error bounds for such realistic systems in the published literature, but they are for estimators that are computationally intractable. In this paper, we develop error bounds for a computationally tractable estimator which also applies to realistic compressive systems obeying the required constraints. The focus of our technique is on the replacement of the generalized Kullback-Leibler divergence, with an information theoretic metric - namely the square root of the Jensen-Shannon divergence, which is related to an approximate, symmetrized version of the Poisson log likelihood function. We show that this replacement allows for very simple proofs of the error bounds, as it proposes and proves several interesting statistical properties of the square root of Jensen-Shannon divergence, and exploits other known ones. Numerical experiments are performed showing the practical use of the technique in signal and image reconstruction from compressed measurements under Poisson noise. Our technique is applicable to signals that are sparse or compressible in any orthonormal basis, works with high probability for any randomly generated sensing matrix that obeys the non-negativity and flux preservation constraints, and is based on an estimator whose parameters are purely statistically motivated. 
\end{abstract}

\end{frontmatter}

\section{Introduction}
{C}{ompressed} sensing is today a very mature field of research in signal processing, with several advances on the theoretical, algorithmic as well as application fronts. The theory essentially considers measurements of the form $\boldsymbol{y} = \boldsymbol{\Phi} \boldsymbol{x} = \boldsymbol{\Phi \Psi \theta} = \boldsymbol{A \theta}$ where $\boldsymbol{y} \in \mathbb{R}^N$ is a measurement vector, $\boldsymbol{A} \in \mathbb{R}^{N \times m}$ is the product of a sensing matrix $\boldsymbol{\Phi}$ (with much fewer rows than columns, \textit{i.e.}, $N \ll m$), $\boldsymbol{\Psi} \in \mathbb{R}^{m \times m}$ is a signal representation orthonormal basis, and $\boldsymbol{\theta} \in \mathbb{R}^m$ is a vector that is sparse or compressible such that $\boldsymbol{x} = \boldsymbol{\Psi \theta}$. Under suitable conditions on the sensing matrix such as the restricted isometry property (RIP) and sparsity-dependent lower bounds on $N$, it is proved that $\boldsymbol{x}$ can be recovered near-accurately given $\boldsymbol{y}$ and $\boldsymbol{\Phi}$, even if the measurement $\boldsymbol{y}$ is corrupted by signal-independent, additive noise $\boldsymbol{\eta}$ of the form $\boldsymbol{y} = \boldsymbol{\Phi x} + \boldsymbol{\eta}$ where $\boldsymbol{\eta} \sim \mathcal{N}(0,\sigma^2)$ or $\|\boldsymbol{\eta}\|_2 \leq \epsilon$ (bounded noise). The specific error bound \cite{Candes2008} on $\boldsymbol{\theta}$ in the case of $\|\boldsymbol{\eta}\|_2 \leq \epsilon$ is given as:
\begin{equation}
\|\boldsymbol{\theta}-\boldsymbol{\theta^\star}\|_2 \leq C_1 \epsilon + \dfrac{C_2}{\sqrt{s}} \|\boldsymbol{\theta}-\boldsymbol{\theta_s}\|_1
\label{eq:CS1}
\end{equation}
where $\boldsymbol{\theta_s}$ is a vector created by setting all entries of $\boldsymbol{\theta}$ to 0 except for those containing the $s$ largest absolute values, $\boldsymbol{\theta^\star}$ is the minimum of the following optimization problem denoted as (\textsf{P1}),
\begin{equation}
\textrm{(\textsf{P1}): minimize} \|\boldsymbol{z}\|_1 \textrm{ such that } \|\boldsymbol{y}-\boldsymbol{Az}\|_2 \leq \epsilon,
\label{eq:l1_l2_opt}
\end{equation}
and $C_1$ and $C_2$ are constants independent of $m$ or $N$ but dependent only on $\delta_{2s}$, the so-called restricted isometry constant (RIC) of $\boldsymbol{A}$. These bounds implicity require that $N \sim \Omega(s \log m)$.

The noise affecting several different types of imaging systems is, however, known to follow the Poisson distribution. Examples include photon-limited imaging systems deployed in night-time photography \cite{Alter2006}, astronomy \cite{Starck2010}, low-dosage CT or X-ray imaging \cite{Boone2003} or fluorescence microscopy \cite{Boulanger2010, SYang2015}. The Poisson noise model is given as follows:
\begin{equation}
\boldsymbol{y} \sim \textrm{Poisson}(\boldsymbol{\Phi x})
\end{equation}
where $\boldsymbol{x} \in \mathbb{R}_{\geq 0}^m$ is the \emph{non-negative} signal or image of interest. The likelihood of observing a given measurement vector $\boldsymbol{y}$ is given as 
\begin{equation}
p(\boldsymbol{y}|\boldsymbol{\Phi x}) = \prod_{i=1}^n \dfrac{[(\boldsymbol{\Phi x})_i]^{y_i} e^{-(\boldsymbol{\Phi x})_i}}{y_i!}
\end{equation}
where $y_i$ and $(\boldsymbol{\Phi x})_i$ are the $i^{\textrm{th}}$ component of the vectors $\boldsymbol{y}$ and $\boldsymbol{\Phi x}$ respectively. 

Unfortunately, the mathematical guarantees for compressive reconstruction from bounded or Gaussian noise \cite{Cai2013,Candes2008,Studer2012} are no longer \emph{directly} applicable to the case where the measurement noise follows a Poisson distribution, which is the case considered in this paper. One important reason for this is a feature of the Poisson distribution - that the mean and the variance are equal to the underlying intensity, thus deviating from the signal independent or bounded nature of other noise models. 

Furthermore, the aforementioned practical imaging systems essentially act as photon-counting systems. Not only does this require non-negative signals of interest, but it also imposes crucial constraints on the nature of the sensing matrix $\boldsymbol{\Phi}$:
\begin{enumerate}
\item Non-negativity: $\forall i, \forall j, {\Phi}_{ij} \geq 0$
\item Flux-preservation: The total photon-count of the observed signal $\boldsymbol{\Phi x}$ can never exceed the photon count of the original signal $\boldsymbol{x}$, \textit{i.e.}, $\sum_{i=1}^N (\boldsymbol{\Phi x})_i \leq \sum_{i=1}^m x_i$. This in turn imposes the constraint that every column of $\boldsymbol{\Phi}$ must sum up to a value no more than 1, i.e. $\forall j, \sum_{i=1}^N {\Phi}_{ij} \leq 1$.
\end{enumerate}
A randomly generated non-negative and flux-preserving $\boldsymbol{\Phi}$ matrix does \emph{not} (in general) obey the RIP. This situation is in contrast to randomly generated Gaussian or Bernoulli ($\pm 1$) random matrices which obey the RIP with high probability \cite{Baraniuk2008}, and poses several challenges. However following prior work \cite{Raginsky2010}, we construct a related matrix $\boldsymbol{\tilde{\Phi}}$ from $\boldsymbol{\Phi}$ which obeys the RIP.

\subsection{Main Contributions}
\label{subsec:main_c}
The derivation of the theoretical performance bounds in Eqn. \ref{eq:CS1} based on the optimization problem in Eqn. \ref{eq:l1_l2_opt} cannot be used in the Poisson noise model case, as it is well known that the use of the $\ell_2$ norm between $\boldsymbol{y}$ and $\boldsymbol{\Phi x}$ leads to oversmoothing in the lower intensity regions and undersmoothing in the higher intensity regions. To estimate an unknown parameter set $\boldsymbol{x}$ given a set of Poisson-corrupted measurements $\boldsymbol{y}$, one proceeds by the maximum likelihood method. Dropping terms involving only $\boldsymbol{y}$, this reduces to maximization of the quantity  $\sum_{i=1}^N y_i \log \dfrac{y_i}{(\boldsymbol{\Phi x})_i} - \sum_{i=1}^n y_i + \sum_{i=1}^n (\boldsymbol{\Phi x})_i$ which is called the generalized Kullback-Leibler divergence \cite{Fevotte2009} between $\boldsymbol{y}$ and $\boldsymbol{\Phi x}$ - denoted as $G(\boldsymbol{y},\boldsymbol{\Phi x})$. This divergence measure, however, does not obey the triangle inequality, quite unlike the $\ell_2$ norm term in Eqn. \ref{eq:l1_l2_opt} which is a metric. This `metric-ness' of the $\ell_2$ norm constraint is an important requirement for the error bounds in Eqn. \ref{eq:CS1} proved in \cite{Candes2008}. For instance, the triangle inequality of the $\ell_2$ norm is used to prove that $\|\boldsymbol{A}(\boldsymbol{\theta}-\boldsymbol{\theta^\star})\|_2 \leq 2\epsilon$ where $\boldsymbol{\theta^\star}$ is the minimizer of Problem (\textsf{P1}) in Eqn. \ref{eq:l1_l2_opt}. This is done in the following manner:
\begin{eqnarray}
\|\boldsymbol{A}(\boldsymbol{\theta}-\boldsymbol{\theta^\star})\|_2 \leq \|\boldsymbol{y}-\boldsymbol{A}\boldsymbol{\theta}\|_2 + \|\boldsymbol{y}-\boldsymbol{A}\boldsymbol{\theta^\star}\|_2 \leq 2\epsilon.
\end{eqnarray}
This upper bound on $\|\boldsymbol{A}(\boldsymbol{\theta}-\boldsymbol{\theta^\star})\|_2$ is a crucial step in \cite{Candes2008}, for deriving the error bounds of the form in Eqn. \ref{eq:CS1}.

The $\ell_2$ norm is however not appropriate for the Poisson noise model for the aforementioned reasons. The first major contribution of this paper is to replace the $\ell_2$ norm error term by a term which is more appropriate for the Poisson noise model and which, at the same time, is a metric. The specific error term that we choose here is the square root of the Jensen-Shannon divergence, which is a well-known information theoretic metric \cite{Endres2003}. Hereafter we abbreviate the Jensen-Shannon divergence as JSD, its square-root as SQJSD, and denote them as $J$ and $\sqrt{J}$ respectively within equations. Let $\boldsymbol{\theta^\star}$ be the minimizer of the following optimization problem which we denote as (\textsf{P2}):
\begin{eqnarray}\label{eq:l1_sqjsd}
\textrm{(\textsf{P2}): minimize} \|\boldsymbol{z}\|_1 \textrm{ such that } \sqrt{J(\boldsymbol{y},\boldsymbol{Az})} \leq \epsilon, \boldsymbol{z} \succeq \boldsymbol{0}, \\ \nonumber
\|\boldsymbol{\Psi z}\|_1 = I,
\end{eqnarray}
where $I \triangleq \sum_{i=1}^m x_i$ is the total intensity of the signal of interest and $\epsilon$ is an upper bound on $\sqrt{J(\boldsymbol{y},\boldsymbol{Az})}$ that we set to $\sqrt{N}(\frac{1}{2}+\frac{\sqrt{11}}{8})$ (for reasons that will be clear in Section \ref{sec:Main_result} and \ref{sec:proofs}). We then prove that with high probability
\begin{equation}
\dfrac{\|\boldsymbol{\theta}-\boldsymbol{\theta^\star}\|_2}{I} \leq C_1 \mathcal{O}\bigg( \dfrac{N}{\sqrt{I}}\bigg) + \dfrac{C_2}{I\sqrt{s}} \|\boldsymbol{\theta}-\boldsymbol{\theta_s}\|_1
\label{eq:CS2}
\end{equation}
where $C_1$ and $C_2$ are constants that depend only on the RIC of the sensing matrix $\boldsymbol{\tilde{\Phi}}$ derived from $\boldsymbol{\Phi}$. This result is proved in Section \ref{sec:Main_result}, followed by an extensive discussion. In particular, we explain the reason behind the apparently counter-intuitive first term which is increasing in $N$: namely, that a Poisson imaging system distributes the total incident photon flux across the $N$ measurements, reducing the SNR per measurement and hence affecting the performance. This phenomenon has been earlier observed in \cite{Raginsky2010}. Our performance bounds derived independently and via a completely different method confirm the same phenomenon.

While there exists a body of earlier work on reconstruction error bounds for Poisson corrupted compressive measurements \cite{Raginsky2010,Jiang2015,Rish2009,Ivanoff2016,Jiang2015_arxiv,Rohban2016}, the approach taken in this paper is different, and has the following features:
\begin{enumerate}
\item Existing techniques such as \cite{Raginsky2010,Jiang2015} work with intractable estimators for Poisson compressed sensing although they are designed to deal with physically realizable compressive systems. On the other hand, there are several techniques such as \cite{Rish2009,Ivanoff2016,Jiang2015_arxiv,Rohban2016} which are applicable to computationally efficient estimators (convex programs) and produce provable guarantees, but they do not impose important constraints required for physical implementability.  Our approach, however, works with a computationally tractable estimator involving regularization with the $\ell_1$ norm of the sparse coefficients representing the signal, while at the same time being applicable to physically realizable compressive systems. See Section \ref{sec:prior_work} for a detailed comparison.
\item Our technique demonstrates successfully (for the first time, to the best of our knowledge) the use of the JSD and the SQJSD for Poisson compressed sensing problems, at a theoretical as well as experimental level. Our work exploits several interesting properties of the JSD, some of which we derive in this paper. Our suggested numerical procedure does not require tweaking of a regularization parameter, but uses a constrained optimization procedure with a parameter dictated by the statistical properties of the SQJSD as shown in Section \ref{subsec:JSD_SQJSD}.
\item Our technique affords (arguably) much simpler proofs than existing methods.
\end{enumerate}

\subsection{Organization of the Paper}
The main theoretical result is derived in detail in Section \ref{sec:Main_result}, especially Section \ref{subsec:JSD_SQJSD}. Numerical simulations are presented in Section \ref{sec:numericals}. Relation to prior work on Poisson compressed sensing is examined in detail in Section \ref{sec:prior_work}, followed by a discussion in Section \ref{sec:conclusion}. The proofs of some key theorems are presented in Section \ref{sec:proofs}. The relation between the JSD and a symmetrized version of the Poisson likelihood is examined in Section \ref{sec:relation}.

\section{Main Result}
\label{sec:Main_result}
\subsection{Construction of Sensing Matrices}
\label{subsec:sm}
We construct a sensing matrix $\boldsymbol{\Phi}$ ensuring that it corresponds to the forward model of a real optical system, based on the approach in \cite{Raginsky2010}. Therefore it has to satisfy certain properties imposed by constraints of a physically realizable optical system - namely non-negativity and flux preservation. One major difference between Poisson compressed sensing and conventional compressed sensing emerges from the fact that conventional randomly generated sensing matrices which obey RIP do not follow the aforementioned physical constraints (although sensing matrices can be \emph{designed} to obey the RIP, non-negativity and flux preservation simultaneously as in \cite{Mordechay2014}, and we comment upon this aspect in the remarks following the proof of our key theorem, later on in this section). In the following, we construct a sensing matrix $\boldsymbol{\Phi}$ which has only zero or (scaled) ones as entries. Let us define $p$ to be the probability that a matrix entry is 0, then $1-p$ is the probability that the matrix entry is a scaled 1. Let $\boldsymbol{Z}$ be a $N \times m$ matrix whose entries $Z_{i,j}$ are i.i.d random variables defined as follows,
\begin{subnumcases}{Z_{i,j}=}
-\sqrt{\dfrac{1-p}{p}} & with probability $p$,
\\
\sqrt{\dfrac{p}{1-p}} & with probability $1-p$.
\end{subnumcases}
Let us define $\boldsymbol{\tilde{\Phi}} \triangleq \dfrac{\boldsymbol{Z}}{\sqrt{N}}$. For $p = 1/2$, the matrix $\boldsymbol{\tilde{\Phi}}$ now follows RIP of order $2s$ with a very high probability given by $1-2e^{-Nc(1+\delta_{2s})}$ where $\delta_{2s}$ is its RIC of order $2s$ and function $c(h) \triangleq \dfrac{h^2}{4}-\dfrac{h^3}{6}$ \cite{Baraniuk2008}. In other words, for any $2s$-sparse signal $\boldsymbol{\rho}$, the following holds with high probability
\begin{equation*} (1-\delta_{2s})\|\boldsymbol{\rho}\|^2_2 \leq \|\boldsymbol{\tilde{\Phi}\rho}\|^2_2 \leq (1+\delta_{2s})\|\boldsymbol{\rho}\|^2_2. \end{equation*} 
Given any orthonormal matrix $\boldsymbol{\Psi}$, arguments in \cite{Baraniuk2008} show that $\boldsymbol{\tilde{\Phi}\Psi}$ also obeys the RIP of the same order as $\boldsymbol{\tilde{\Phi}}$. 

However $\boldsymbol{\tilde{\Phi}}$ will clearly contain negative entries with very high probability, which violates the constraints of a physically realizable system. To deal with this, we construct the flux-preserving and positivity preserving sensing matrix $\boldsymbol{\Phi}$ from $\boldsymbol{\tilde{\Phi}}$ as follows:
\begin{equation} 
\boldsymbol{\Phi} \triangleq \sqrt{\dfrac{p(1-p)}{N}} \boldsymbol{\tilde{\Phi}} + \dfrac{(1-p)}{N}\boldsymbol{1}_{N \times m}, 
\label{eq:Phi} 
\end{equation}
which ensures that each entry of $\boldsymbol{\Phi}$ is either $0$ or $\dfrac{1}{N}$. In addition, one can easily check that $\boldsymbol{\Phi}$ satisfies both the non-negativity as well as flux-preservation properties. 

\subsection{The Jensen-Shannon Divergence and its Square Root}
\label{subsec:JSD_SQJSD}
The well-known Kullback-Leibler Divergence between vectors $\boldsymbol{p} \in {\mathbb{R}_{\geq 0}}^{n \times 1}$ and $\boldsymbol{q} \in {\mathbb{R}_{\geq 0}}^{n \times 1}$ denoted by $D(\boldsymbol{p},\boldsymbol{q})$ is defined as\footnote{Note that the Kullback-Leibler and other divergences are usually defined for probability mass functions, but they have also been used in the context of general non-negative vectors in the same manner as we do in this paper.} 
\begin{equation} D(\boldsymbol{p},\boldsymbol{q}) \triangleq \sum_{i=1}^{n} p_i\log{\dfrac{p_i}{q_i}}.\label{eq:D}\end{equation}
The Jensen-Shannon Divergence between $\boldsymbol{p}$ and $\boldsymbol{q}$ denoted by $J(\boldsymbol{p},\boldsymbol{q})$ is defined as
\begin{equation}J(\boldsymbol{p},\boldsymbol{q}) \triangleq \dfrac{D(\boldsymbol{p},\boldsymbol{m}) + D(\boldsymbol{q},\boldsymbol{m})}{2}\end{equation}
where $\boldsymbol{m} \triangleq \dfrac{1}{2}(\boldsymbol{p}+\boldsymbol{q})$.

\begin{figure*}[!t]
\centering
\includegraphics[width=2.3in]{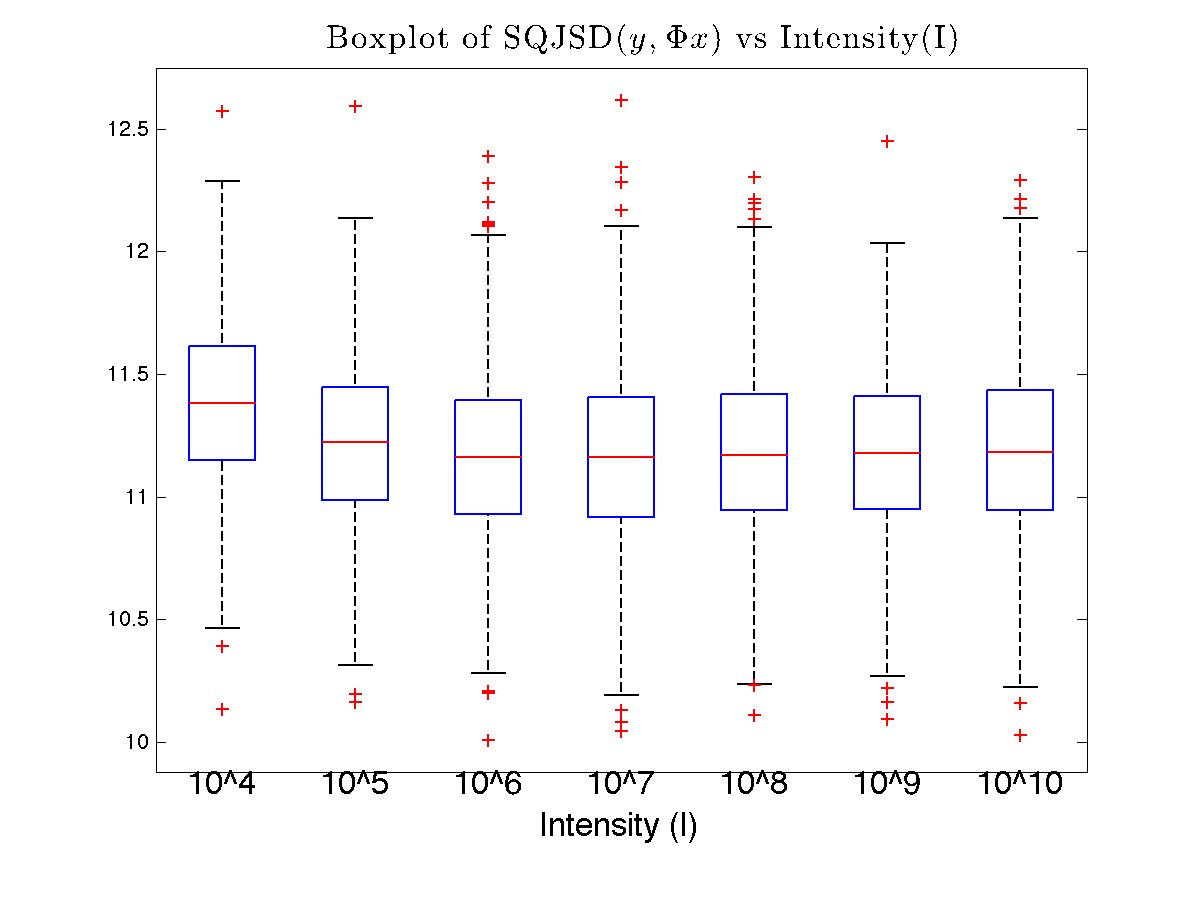}
\includegraphics[width=2.3in]{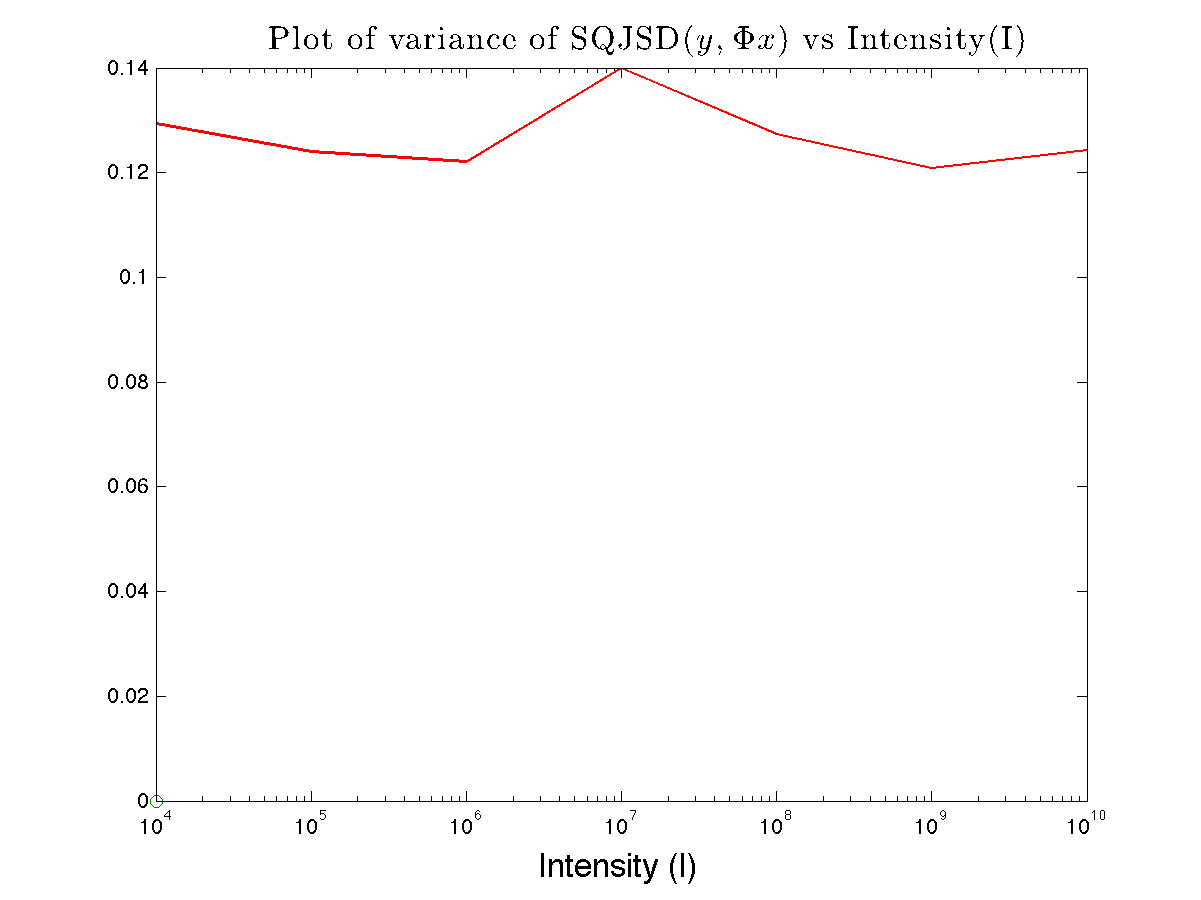}
\includegraphics[width=2.3in]{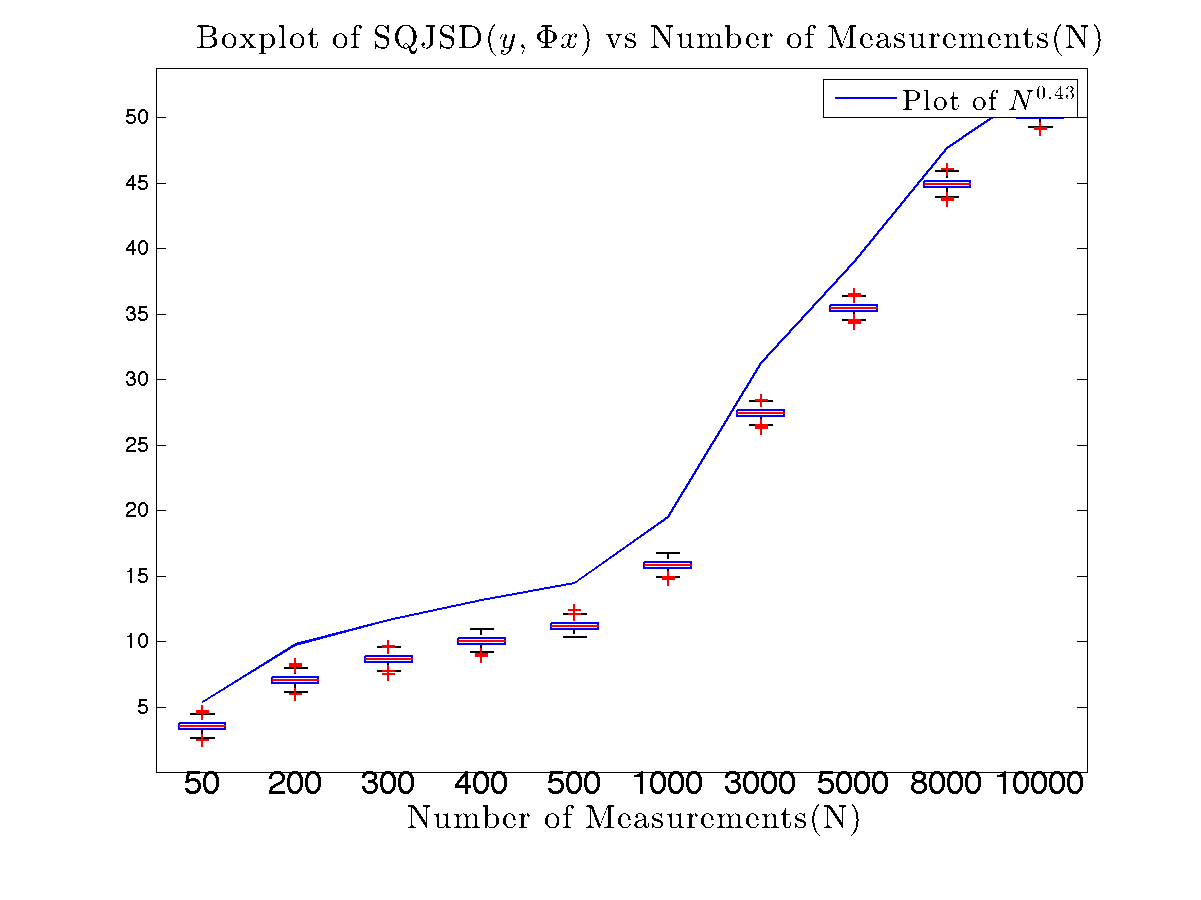}
\includegraphics[width=2.3in]{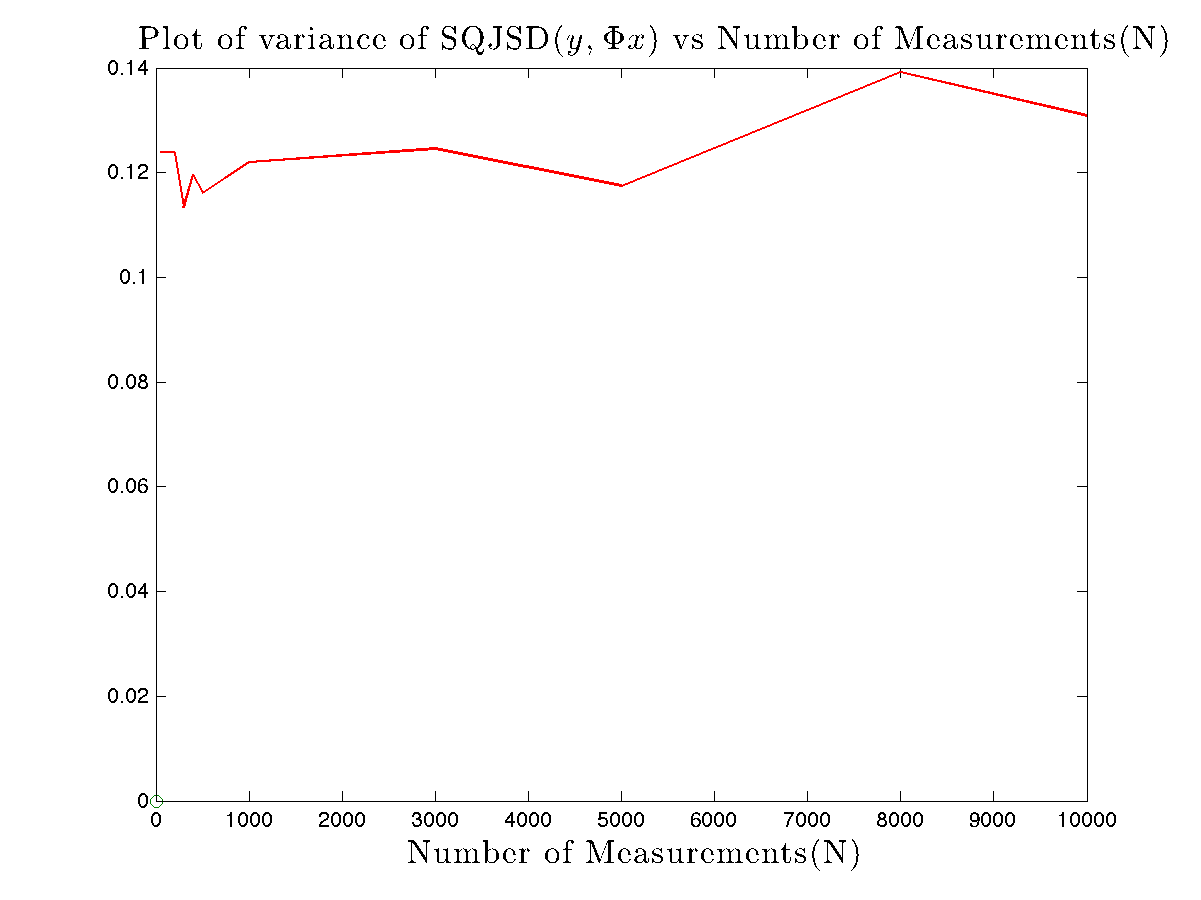}
\includegraphics[width=3in]{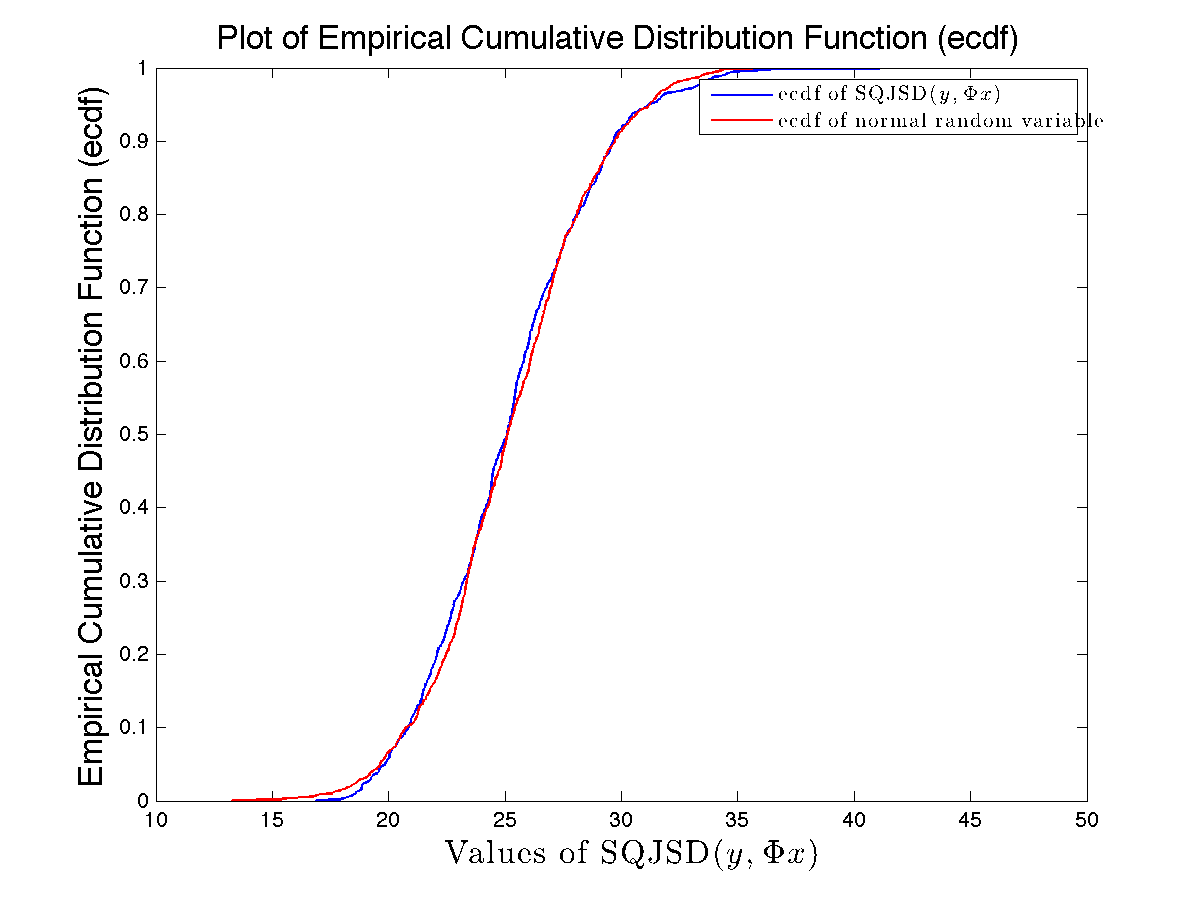} 
\caption{First row: Box plot and plot of variance of the values of $\sqrt{J(\boldsymbol{y},\boldsymbol{\Phi x})}$ versus $I$ for a fixed $N = 500$ for a signal of dimension $m = 1000$.
Second row: Box plot and plot of the variance of the values of $\sqrt{J(\boldsymbol{y},\boldsymbol{\Phi x})}$ versus $N$ for a fixed $I = 10^6$ for a signal of dimension $m = 1000$. The line above the box-plots in the top figure represents the curve for $N^{0.43}$. Third row: Empirical CDF of $\sqrt{J(\boldsymbol{y},\boldsymbol{\Phi x})}$ for $N = 100, I = 10^4, m = 500$ compared to a Gaussian CDF with mean and variance equal to that of the values of $\sqrt{J(\boldsymbol{y},\boldsymbol{\Phi x})}$.
Scripts for reproducing all results at \cite{suppcode}.}
\label{fig:SQJSD_prop}
\end{figure*}

The performance bounds derived in this paper for reconstruction from Poisson-corrupted measurements deal with the estimate obtained by solving the constrained optimization problem (\textsf{P2}) in Eqn. \ref{eq:l1_sqjsd}, where we consider an upper bound of $\epsilon$ on the SQJSD. The motivation for this formulation will be evident from the following properties of the JSD considered in this section: (1) the metric nature of (including the triangle inequality observed by) its square-root,  (2) its relation with the total variation distance $V(\boldsymbol{p},\boldsymbol{q}) \triangleq \sum_i |p_i-q_i|$, and (3) interesting statistical properties of $\sqrt{J(\boldsymbol{y},\boldsymbol{\Phi x})}$. These properties, the last of which are proved in this paper, are very useful in deriving the performance bounds in the following sub-section. \\
\textbf{Lemma 1:} The square root of the Jensen-Shannon Divergence is a metric \cite{Endres2003}.\\
\textbf{Lemma 2:} Let us define
\begin{equation*} \begin{split}
V(\boldsymbol{p},\boldsymbol{q}) \triangleq \sum_{i=1}^{n} |{p_i - q_i}| \\
\Delta(\boldsymbol{p},\boldsymbol{q})  \triangleq \sum_{i=1}^n {\dfrac{{|p_i - q_i|}^2}{p_i + q_i}}. \nonumber
\end{split} \end{equation*}
If $\boldsymbol{p}, \boldsymbol{q} \succeq \boldsymbol{0}$ and $\|\boldsymbol{p}\|_1 \leq 1$, $\|\boldsymbol{q}\|_1 \leq 1$ then as per \cite{Endres2003},
\begin{equation}  \dfrac{1}{2}V(\boldsymbol{p},\boldsymbol{q})^2 \leq \Delta(\boldsymbol{p},\boldsymbol{q})  \leq 4 J(\boldsymbol{p},\boldsymbol{q}).\end{equation}
Additionally, \emph{we have experimentally observed some interesting properties of the distribution of the SQJSD values}, across different Poisson realizations of compressive measurements of a signal $\boldsymbol{x}$, acquired with a fixed and realistic sensing matrix $\boldsymbol{\Phi}$ as described in Section \ref{subsec:sm}. In other words, if $\boldsymbol{y} \sim \textrm{Poisson}(\boldsymbol{\Phi x})$, then we consider the distribution of $\sqrt{J(\boldsymbol{y},\boldsymbol{\Phi x})}$ across different realizations of $\boldsymbol{y}$. Our observations, shown in Figure \ref{fig:SQJSD_prop} are as follows:
\begin{enumerate}
\item Beyond a threshold $\tau$ on the intensity $I$, the expected value of $\sqrt{J(\boldsymbol{y},\boldsymbol{\Phi x})}$ is nearly constant (say some $\kappa$), and independent of $I$, given a fixed number of measurements $N$. For $I \leq \tau$, we have $\sqrt{J(\boldsymbol{y},\boldsymbol{\Phi x})} \leq \kappa$. 
\item The variance $\sigma^2$ of $\sqrt{J(\boldsymbol{y},\boldsymbol{\Phi x})}$ is small, irrespective of the value of $I$ and $N$. 
\item For any $I$, the mean (and any chosen percentile, such as the 99 percentile) of $\sqrt{J(\boldsymbol{y},\boldsymbol{\Phi x})}$ scales as $O(N^{0.5})$ w.r.t. $N$ with a constant factor very close to 1. 
\item Irrespective of $I$, $N$ or $m$, the distribution of $\sqrt{J(\boldsymbol{y},\boldsymbol{\Phi x})}$ is Gaussian with mean and standard deviation equal to the empirical mean and empirical standard deviation of the values of $\sqrt{J(\boldsymbol{y},\boldsymbol{\Phi x})}$. This is confirmed by a Kolmogorov-Smirnov (KS) test even at 1\% significance (see \cite{suppcode}). 
\end{enumerate}
We emphasize that as per our extensive simulations, these properties are independent of specific realizations of $\boldsymbol{\Phi}, \boldsymbol{x}$ or the dimensionality or sparsity of $\boldsymbol{x}$. Our scripts to reproduce these results are included at \cite{suppcode}. Our attempt to formalize these observations lead to the following theorem which we prove in Section \ref{sec:proofs}.\\
\textbf{Theorem 1:} Let $\boldsymbol{y} \in \mathbb{Z}^N_{+}$ be a vector of compressive measurements such that $y_i \sim \textrm{Poisson}[(\boldsymbol{\Phi x})_i]$ where $\boldsymbol{\Phi} \in \mathbb{R}^{N \times m}$ is a non-negative flux-preserving matrix and $\boldsymbol{x} \in \mathbb{R}^m$ is a non-negative signal. Define $s_i \triangleq N \times (\boldsymbol{\Phi x})_i$. Then we have:
\begin{enumerate}
\item $E[\sqrt{J(\* y, \*\Phi\*x)}] \leq \sqrt{N/4}$
\item $v \triangleq \textrm{Var}[\sqrt{J(\* y, \*\Phi\*x)}] \leq \dfrac{11+5 \sum_{i=1}^N 1/s_i}{\textrm{max}(0,4(2-\sum_{i=1}^N 1/s_i))}$
\item $P\Big(\sqrt{J(\* y, \*\Phi\*x)} \leq \sqrt{N}(\frac{1}{2}+\frac{\sqrt{11}}{8})\Big) \geq 1-2e^{-N/2}$ for some constant $d$.
\end{enumerate}
We make a few comments below:
\begin{enumerate}
\item $E[\sqrt{J(\* y, \*\Phi\*x)}]$ does not increase with $I$. This property is \emph{not} shared by the negative log-likelihood of the Poisson distribution. This forms one major reason for using SQJSD as opposed to the latter, for deriving the bounds in this paper.
\item If each $s_i$ is sufficiently large in value (\textit{i.e.} $\gg 0.5$), this yields $\textrm{Var}[\sqrt{J(\* y, \*\Phi\*x)}] \lessapprox \frac{11}{8}$ which is independent of $N$ as well as the measurement or signal values. See also the simulation in Figure \ref{fig:SQJSD_prop}.
\item The assumption that $s_i \gg 0.5$ is not restrictive in most signal or image processing applications, except those that work with extremely low intensity levels. In the latter case, our variance bound is less useful. But in such cases the performance of Poisson compressed sensing is itself very poor due to the very low SNR \cite{Jiang2015}. 
\item The last statement of this theorem is based on the central limit theorem, and hence for a finite value of $N$, it is an approximation. However, the approximation is empirically observed to be tight even for small $N \sim 10$ as confirmed by a Kolmogorov-Smirnov test (see \cite{suppcode}). 
\end{enumerate}

\subsection{Theorem on Reconstruction Error Bounds}
\textbf{Theorem 2:} Consider a non-negative signal of interest $\boldsymbol{x} = \boldsymbol{\Psi \theta}$ for orthonormal basis $\boldsymbol{\Psi}$ with sparse vector $\boldsymbol{\theta}$. Define $\boldsymbol{A} \triangleq \boldsymbol{\Phi \Psi}$ for sensing matrix $\boldsymbol{\Phi}$ defined in Eqn. \ref{eq:Phi}. Suppose $\boldsymbol{y}  \sim \textrm{Poisson}(\boldsymbol{\Phi \Psi \theta})$, \textit{i.e.}  $\boldsymbol{y}  \sim \textrm{Poisson}(\boldsymbol{A \theta})$, represents a vector of $N \ll m$ independent Poisson-corrupted compressive measurements of $\boldsymbol{x}$, \textit{i.e.}, $\forall i, 1 \leq i \leq N, y_i \sim \textrm{Poisson}((\boldsymbol{A \theta})_i)$. Let $\boldsymbol{\theta^\star}$ be the solution to the problem (\textsf{P2}) defined earlier, with the upper bound $\epsilon$ in (\textsf{P2}) set to $\sqrt{N}\Big(\frac{1}{2}+\frac{\sqrt{11}}{8}\Big)$. If $\boldsymbol{\tilde{\Phi}}$ constructed from $\boldsymbol{\Phi}$ obeys the RIP of order $2s$ with RIC $\delta_{2s} < \sqrt{2}-1$, then we have
\begin{equation} \begin{split}
\textrm{Pr}\Big( \dfrac{\|\boldsymbol{\theta - \theta^{\star}}\|_2}{I}  &\leq  \tilde{C}\dfrac{N}{\sqrt{I}} + \dfrac{C'' s^{-1/2} \|\boldsymbol{\theta}-\boldsymbol{\theta}_s\|_1}{I} \Big) \geq 1-2e^{-N/2},
%\dfrac{\|\boldsymbol{\theta - \theta^{\star}}\|_2}{I}  &\leq  C'\dfrac{\sqrt{N} \epsilon}{\sqrt{I}} + \dfrac{C'' s^{-1/2} \|\boldsymbol{\theta}-\boldsymbol{\theta}_s\|_1}{I} 
\end{split} \end{equation}
where $\tilde{C} \triangleq C'(1/2+\sigma)$, $C' \triangleq \dfrac{4\sqrt{8(1+\delta_{2s})}}{\sqrt{p(1-p)}(1- (1 + \sqrt{2})\delta_{2s})}$, $C'' \triangleq  (\dfrac{2 - 2\delta_{2s} + 2\sqrt{2 \delta_{2s}}}{1- (1 + \sqrt{2})\delta_{2s}})$, $\boldsymbol{\theta}_s$ is a vector containing the $s$ largest absolute value elements from $\boldsymbol{\theta}$, and $\sigma$ is the standard deviation of $\sqrt{J(y_i,(\boldsymbol{\Phi x})_i)}$, which is upper bounded by (approximately) $\frac{\sqrt{11}}{8}$.

Theorem 2 is proved in Section \ref{sec:proofs}. We make several comments on these bounds below. 
\begin{enumerate}
\item Practical implementation of the estimator \textsf{P2} would require supplying a value for $\epsilon$, which is the upper bound on $\sqrt{J(\*y,\*A\*x)}$. This can be provided based on the theoretical analysis of $\sqrt{J(\*y,\*A\*x)}$ from Theorem 1, which motivates the choice $\epsilon = \sqrt{N}\Big(\frac{1}{2}+\sqrt{\frac{11}{64}}\Big)$. In our experiments, we provided a 99 percentile value (see Section \ref{sec:numericals}) which also turns out to be $\mathcal{O}(\sqrt{N})$ and is independent of $\*x$. 
\item We have derived upper bounds on the \emph{relative} reconstruction error, i.e. on $\dfrac{\|\boldsymbol{\theta - \theta^{\star}}\|_2}{I}$ and not on $\|\boldsymbol{\theta - \theta^{\star}}\|_2$. This is because as the mean of the Poisson distribution increases, so does its variance, which would cause an increase in the root mean squared error. But this error would be small in comparison to the average signal intensity. Hence the \emph{relative} reconstruction error is the correct metric to choose in this context. Indeed, $\dfrac{\|\boldsymbol{\theta - \theta^{\star}}\|_2}{I}$ is upper bounded by two terms, both inversely proportional to $I$, reflecting the common knowledge that recontruction under Poisson noise is more challenging if the original signal intensity is lower.
\item The usage of SQJSD, plays a critical role in this proof. First, the term $J$ is related to the Poisson likelihood as will be discussed in Section \ref{sec:relation}. Second, $\sqrt{J}$ is a metric and hence obeys the triangle inequality. Furthermore, $J$ also upper-bounds the total variation norm, as shown in Lemma 2. Both these properties are essential for the derivation of the critical Step 1 - see Section \ref{sec:proofs}.
\item It may seem counter-intuitive that the first error term increases with $N$. However if the original signal intensity remains fixed at $I$, an increase in $N$ simply distributes the photon flux across multiple measurements thereby decreasing the SNR at each measurement and degrading the performance. Similar arguments have been made previously in \cite{Raginsky2010}. This behaviour is a feature of Poisson imaging systems, and is quite different from the Gaussian noise scenario \cite{Candes2007} where the error decreases with increase in $N$ owing to no flux-preservation constraints.
%\item The first error term is directly proportional to $\epsilon$, which is an upper bound on $\sqrt{J(\boldsymbol{y},\boldsymbol{\Phi x})}$, where $\boldsymbol{y}$ is a Poisson corrupted version of $\boldsymbol{\Phi x}$. We have experimentally observed a surprising fact: as $I$ (\textit{i.e.}, the sum total of the values in $\boldsymbol{x}$) increases beyond some small threshold $\tau$, the value of $\sqrt{J(\boldsymbol{y},\boldsymbol{\Phi x})}$ remains almost constant on an average. For $I < \tau$, the values of $\sqrt{J(\boldsymbol{y},\boldsymbol{\Phi x})}$ are lower than this average value. In experiments across signals of varying sparsity and average intensity, with different random $\boldsymbol{\Phi}$ and across several noisy measurements, we have consistently observed this phenomenon. A sample of these results is shown in Figure \ref{fig:epsilon}. This leads us to experimentally conclude that $\epsilon$ does not vary in proportion to $\sqrt{I}$ or $I$. 
\item The above bound holds for a signal sparse/compressible in some orthonormal basis $\boldsymbol{\Psi}$. However, for reconstruction bounds for a non-negative signal sparse/compressible in the \emph{canonical} basis, \textit{i.e.} $\boldsymbol{\Psi} = \boldsymbol{I}$ and hence $\boldsymbol{x} = \boldsymbol{\theta}$, one can solve the following optimization problem which penalizes the $\ell_q$ ($0 < q < 1$) norm instead of the $\ell_1$ norm:
\begin{equation}
\textrm{min}_{\boldsymbol{\theta}} \|\boldsymbol{\theta}\|_q \\
\textrm{ subject to}  \sqrt{\boldsymbol{J}(\boldsymbol{y},\boldsymbol{A\theta})} \leq \epsilon, \\ \nonumber
\|\boldsymbol{\theta}\|_1 = I, \boldsymbol{\theta} \succeq \boldsymbol{0}
\end{equation}
%\begin{figure*}[!t]
%\centering
%\includegraphics[width=6in]{epsilonVsI.png}
%\caption{A box-plot of $J(\boldsymbol{y},I \boldsymbol{\Phi x})$ versus $I$ for a signal $\boldsymbol{x}$ having $\|\boldsymbol{x}\|_1 = 1$. For every $I \in \{10^3,10^4,10^5,10^6,10^7,10^8,10^9,10^{10}\}$, 1000 different Poisson-corrupted samples of $I \boldsymbol{\Phi x}$ were used to generate the box-plot.}
%\label{fig:epsilon}
%\end{figure*}
Performance guarantees for this case can be developed along the lines of the work in \cite{Saab2010}. Other sparsity-promoting terms such as those based on a logarithmic penalty function (which approximates the original $\ell_0$ norm penalty more closely than the $\ell_1$ norm) may also be employed \cite{Candes2008_FAA,Lingenfelter2009}.
\item While imposition of the constraint that $\|\boldsymbol{z}\|_1 = I$ with $I$ being known may appear as a strong assumption, it must be noted that in some compressive camera architectures, it is easy to obtain an estimate of $I$ during acquisition. One example is the Rice Single Pixel Camera \cite{Duarte2008}, where $I$ can be obtained by turning on all the micro-mirrors, thereby allowing the photo-diode to measure the sum total of all values in the signal. The imposition of this constraint has been considered in earlier works on Poisson compressed sensing such as \cite{Raginsky2010} and \cite{Jiang2015}. Furthermore, we note that in our experiments in Section \ref{sec:numericals}, we have obtained excellent reconstructions even without the imposition of this constraint.
\item Measurement matrices in compressed sensing can be specifically designed to have very low coherence, as opposed to the choice of random matrices. Such approaches have been proposed in for a Poisson setting in \cite{Mordechay2014}. Since the coherence value can be used to put an upper bound on the RIC, one can conclude that such matrices will obey RIP even while obeying non-negativity and flux preservation. In case of such matrices which already obey the RIP, the upper bound on the reconstruction error would potentially tighten by a factor of at least $\sqrt{N}$. However, such matrices are obtained as the output of non-convex optimization problems, and there is no guarantee on how low their coherence, and hence their RIC, will be. Indeed, they may not respect the sufficient condition in our proof that $\delta_{2s} < \sqrt{2}-1$.
\end{enumerate}

\section{Numerical Experiments}
\label{sec:numericals}
\begin{figure*}[!t]
\centering
\includegraphics[width=3.5in]{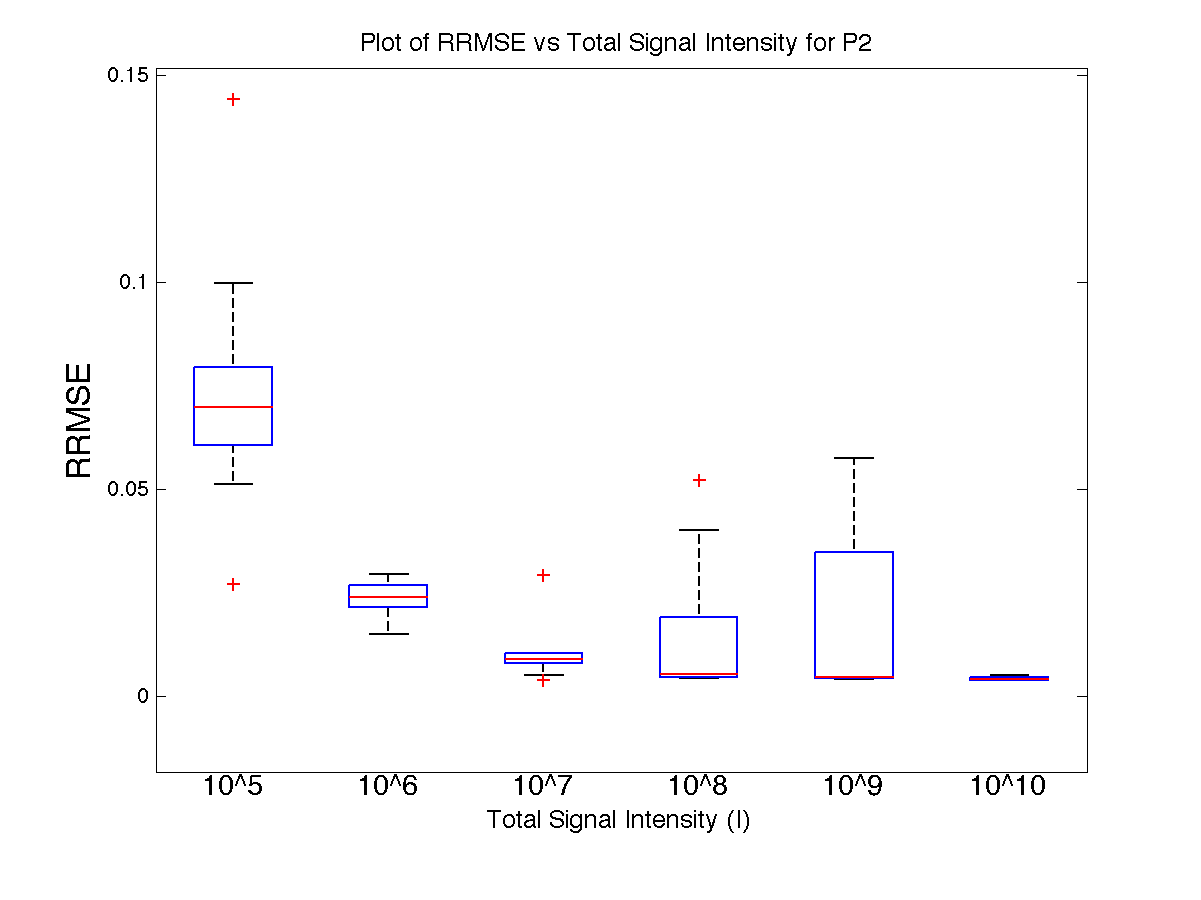}
\includegraphics[width=3.5in]{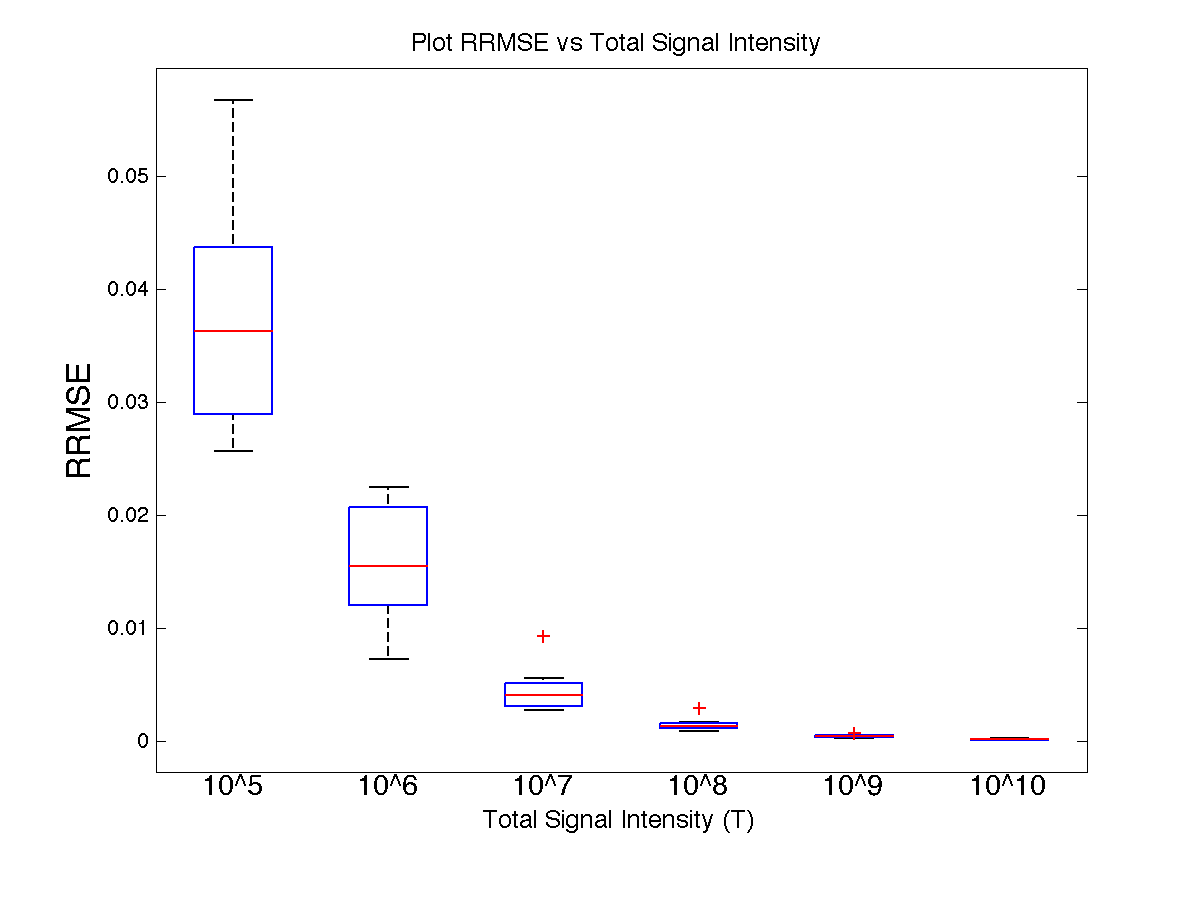}
\caption{Box plots of relative reconstruction error (RRMSE) of problems \textsf{P2} (top) and \textsf{P4} (bottom) for a 1D signal of 100 elements sparse in the canonical basis. RRMSE versus $I$ for a fixed $N = 50$ and fixed sparsity = 5. The $\lambda$ parameter for \textsf{P4} was picked omnisciently (see text for more details).}
\label{fig:graphs}
\end{figure*}

\begin{figure*}[!t]
\centering
\includegraphics[width=3.5in]{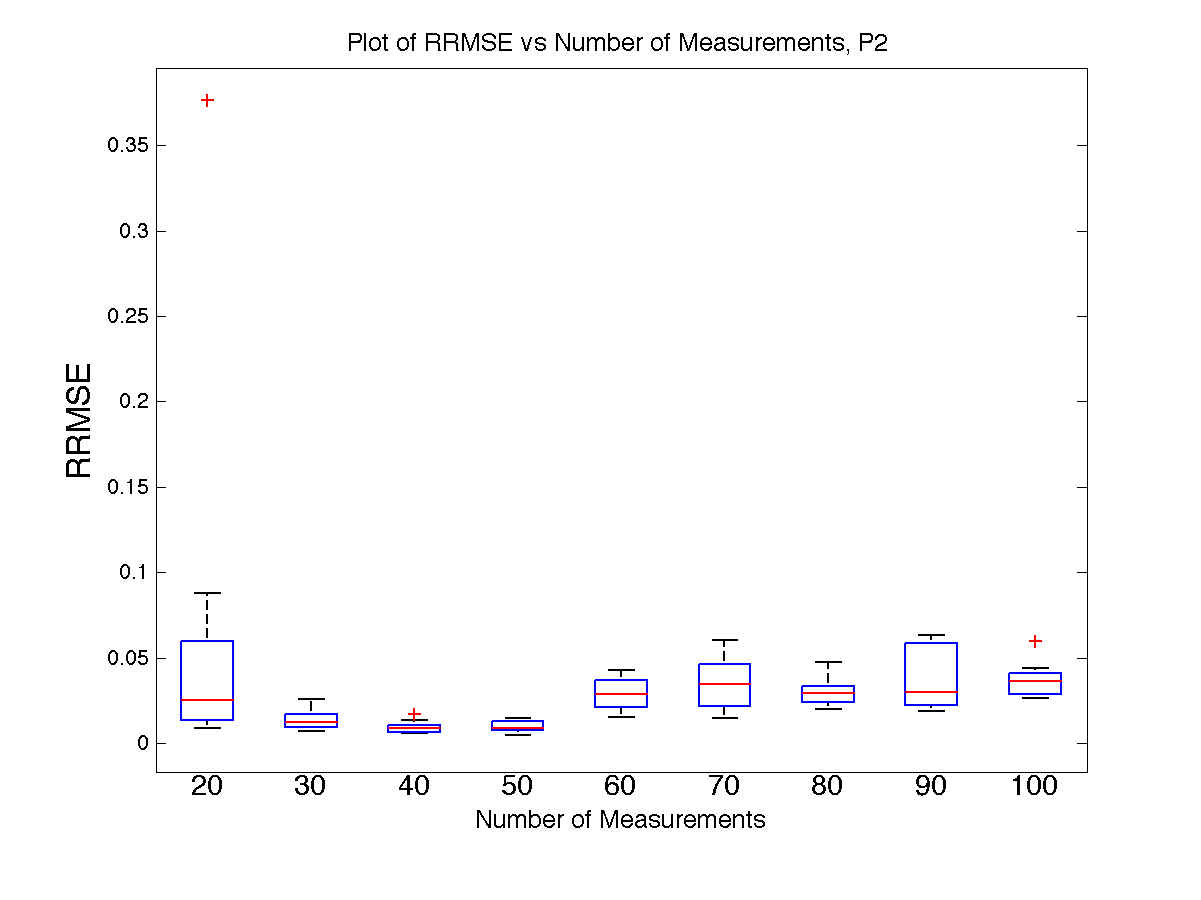}
\includegraphics[width=3.5in]{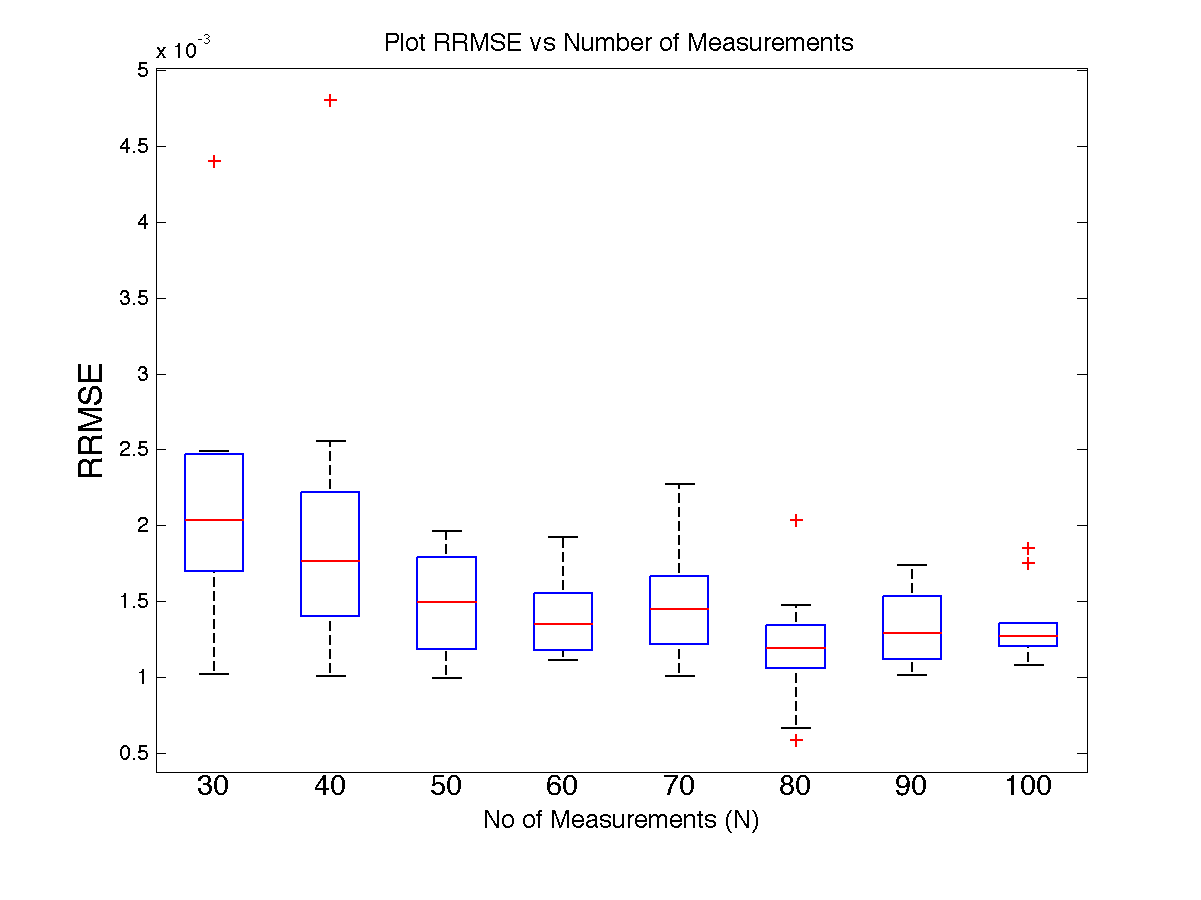}
\caption{Box plots of relative reconstruction error (RRMSE) of problems \textsf{P2} (top) and \textsf{P4} (bottom) for a 1D signal of 100 elements sparse in the canonical basis. RRMSE versus $N$ for a fixed $I = 10^8$ and fixed sparsity = 5. The $\lambda$ parameter for \textsf{P4} was picked omnisciently (see text for more details).}
\label{fig:graphs2}
\end{figure*}

\begin{figure*}[!t]
\centering
\includegraphics[width=3.5in]{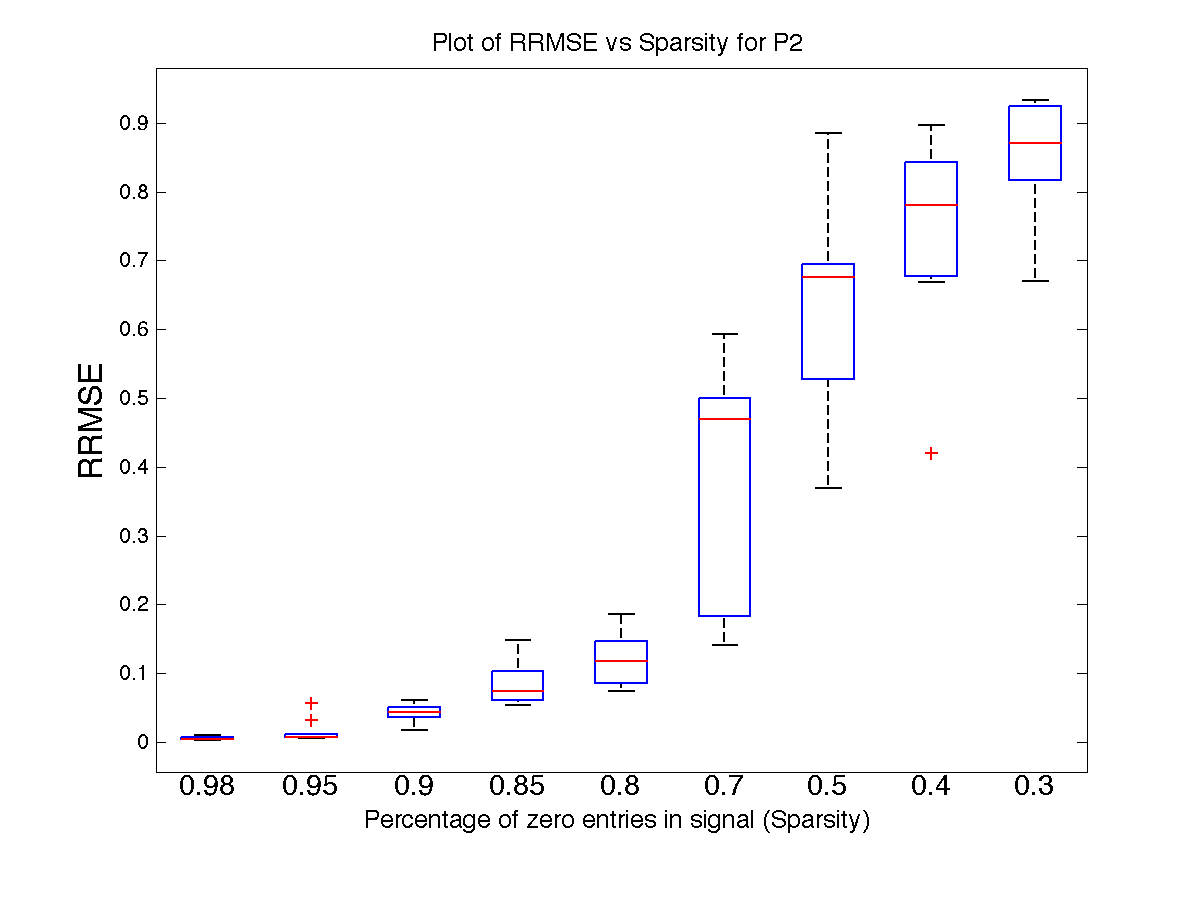}
\includegraphics[width=3.5in]{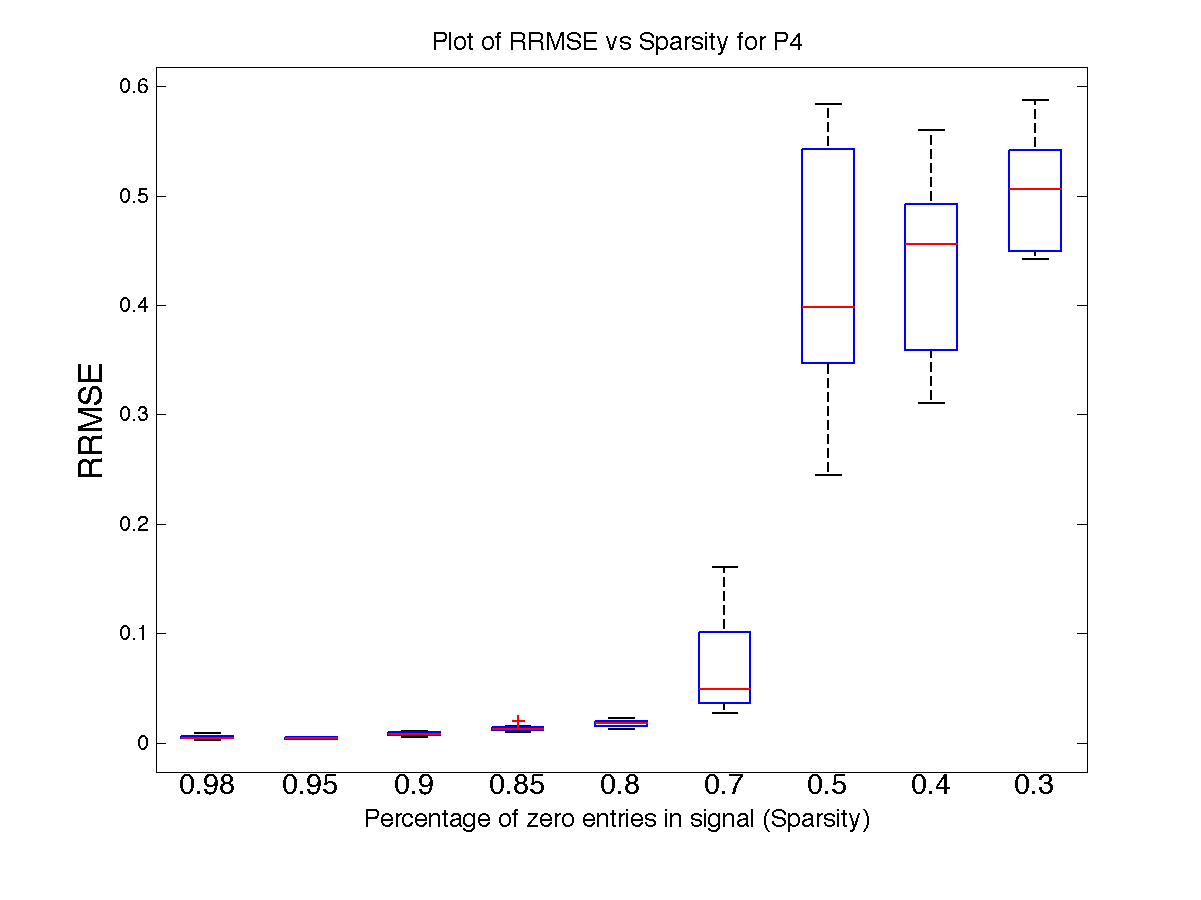}
\caption{Box plots of relative reconstruction error (RRMSE) of problems \textsf{P2} (top) and \textsf{P4} (bottom) for a 1D signal of 100 elements sparse in the canonical basis. RRMSE versus sparsity for a fixed $I = 10^8$ and a fixed $N = 50$. The $\lambda$ parameter for \textsf{P4} was picked omnisciently (see text for more details).}
\label{fig:graphs3}
\end{figure*}

We show results on numerical experiments for problem (\textsf{P2}) without the explicit constraint that $\|\boldsymbol{\Psi \theta}\|_1 = I$, as we obtained excellent results even without it. Besides this, we also show results on the following problem:
\begin{equation}
\textrm{(\textsf{P4}): min} \lambda\|\boldsymbol{\theta}\|_1 + J(\boldsymbol{y},\boldsymbol{\Phi \Psi \theta}) \textrm{ w.r.t. } \boldsymbol{\theta},
\end{equation}
where $\lambda$ is a regularization parameter. Before describing our actual experimental results, we state a lemma that solving (\textsf{P4}) is equivalent to solving (\textsf{P2}) for some pair of $(\lambda,\epsilon)$ values, but again without the constraint $\|\boldsymbol{\Psi \theta}\|_1 = I$. The proof of this lemma follows \cite{Foucart2013} and can be found in the supplemental material in Section \ref{sec:suppmat}.\\
\textbf{Lemma 4:} Given $\boldsymbol{\theta}$ which is the minimizer of problem (\textsf{P4}) for some $\lambda > 0$, there exists some value of $\epsilon = \epsilon_\theta$ for which $\boldsymbol{\theta}$ is the minimizer of problem (\textsf{P2}), but without the constraint $\|\boldsymbol{\Psi \theta}\|_1 = I$.
\\
As JSD is a convex function and $\sqrt{J(\boldsymbol{y},\boldsymbol{\Phi x})} \leq \epsilon$ implies $J(\boldsymbol{y},\boldsymbol{\Phi x}) \leq \epsilon^2$, we solved both (\textsf{P2}) and (\textsf{P4}) using the well-known CVX package \cite{CVX2014} with the SCS solver for native implementation of logarithmic functions\footnote{\url{http://web.cvxr.com/cvx/beta/doc/solver.html}}. The value of $\epsilon$ was chosen to be the 99 percentile of the SQJSD values which are $\mathcal{O}(\sqrt{N})$ and independent of $\*x$ as noted in Section \ref{subsec:JSD_SQJSD}. Experiments were run on Poisson-corrupted compressed measurements obtained from a 1D signal with 100 elements and different levels of sparsity in the canonical (\textit{i.e.}, identity) basis as well as different values of $I$. The sensing matrix followed the architecture discussed in Section \ref{sec:Main_result}. We plotted a graph of the relative reconstruction error given as $RRMSE(\boldsymbol{x},\boldsymbol{x^{\star}}) \triangleq \dfrac{\|\boldsymbol{x}-\boldsymbol{x^{\star}}\|_2}{\|\boldsymbol{x}\|_2}$ versus $I$ for a fixed number of measurements $N = 50$ in Figure \ref{fig:graphs}. This graph clearly reveals lower and lower reconstruction errors with an increase in $I$ which agrees with the worst case error bounds we have derived in this paper. Note that the graph shows box-plots for reconstruction errors for a population of 10 different measurements of a sparse signal using different $\boldsymbol{\Phi}$ matrices. Figure \ref{fig:graphs2} shows a graph with box-plots for $RRMSE(\boldsymbol{x},\boldsymbol{x^{\star}})$ versus $N$ for a fixed $I = 10^8$. Here we observe that the relative error does not decrease significantly with increase in $N$ because of poorer signal to noise ratio with an increase of $N$ and keeping $I$ constant. Lastly, in Figure \ref{fig:graphs3}, we also plotted a graph of average $RRMSE(\boldsymbol{x},\boldsymbol{x^{\star}})$ against signals of different sparsity levels for a fixed $I$ and a fixed $N$. We show comparisons alongside results for problem (\textsf{P4}). While the results of (\textsf{P4}) may appear slightly superior to those of (\textsf{P2}), we emphasize that the parameter $\lambda$ for (\textsf{P4}) was picked \emph{omnisciently}, \textit{i.e.}, assuming the true signal was known and choosing the value of $\lambda$ that gave the least MSE. In practice, this parameter would need to be picked by cross-validation or be a user-choice, whereas there is no such requirement for (\textsf{P2}) since $\epsilon$ is independent of $I$ as shown in Section \ref{subsec:JSD_SQJSD}. 

We also compared our results with the outputs of the following optimization problems:
\begin{eqnarray}
\textrm{(\textsf{P5}): min} \lambda\|\boldsymbol{\theta}\|_1 + \mathcal{SNLL}(\boldsymbol{y},\boldsymbol{\Phi \Psi \theta}) \textrm{ w.r.t. } \boldsymbol{\theta} \\ \nonumber
\textrm{(\textsf{P6}): min} \lambda\|\boldsymbol{\theta}\|_1 + G(\boldsymbol{y},\boldsymbol{\Phi \Psi \theta}) \textrm{ w.r.t. } \boldsymbol{\theta} \nonumber
\end{eqnarray}
since they, especially $G$, can be considered `natural competitors'. Problems (\textsf{P5}) and (\textsf{P6}) were implemented in CVX under the same setting as described for (\textsf{P4}) since $\mathcal{SNLL}$ and $G$ are convex functions. In addition, we also compared these results to those of the well-known Poisson compressed sensing solver known as SPIRAL-TAP from \cite{Harmany2012} which essentially solves (\textsf{P6}) but follows a different optimization method. For (\textsf{P4}), (\textsf{P5}), (\textsf{P6}) and SPIRAL-TAP the regularization parameter $\lambda$ was picked omnisciently (as the results of all these problems were significantly affected by the choice of $\lambda$). We obtained nearly identical results for (\textsf{P4}), (\textsf{P5}), (\textsf{P6}) and SPIRAL-TAP under all settings. Our supplemental material at \url{https://www.cse.iitb.ac.in/~ajitvr/SQJSD/} contains scripts for execution of these results in CVX.

\begin{figure*}[!h]
\includegraphics[width=1.6in]{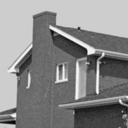}
\includegraphics[width=1.6in]{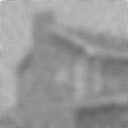}
\includegraphics[width=1.6in]{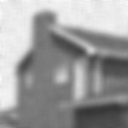}

\includegraphics[width=1.6in]{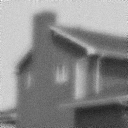}
\includegraphics[width=1.6in]{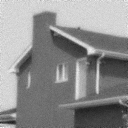}
\includegraphics[width=1.6in]{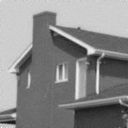}

\includegraphics[width=1.6in]{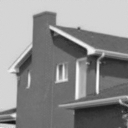}
\includegraphics[width=1.6in]{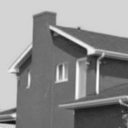}
\caption{Sample reconstruction results for Poisson-corrupted compressed measurements of an image using penalized JSD and a 2D-DCT basis. Left to right, top to bottom: original image, reconstructions for $I = 10^4$, $I = 10^5$, $I = 10^6$, $I = 10^7$, $I = 10^8$, $I = 10^9$, $I = 10^{10}$. The respective relative reconstruction errors (RRMSE) are 0.7, 0.1, 0.0622, 0.03, 0.015, 0.012 and 0.011. Refer to Section \ref{sec:numericals} for more details.}
\label{fig:results_image}
\end{figure*}

\begin{figure*}[!h]
\centering
\includegraphics[width=2.3in]{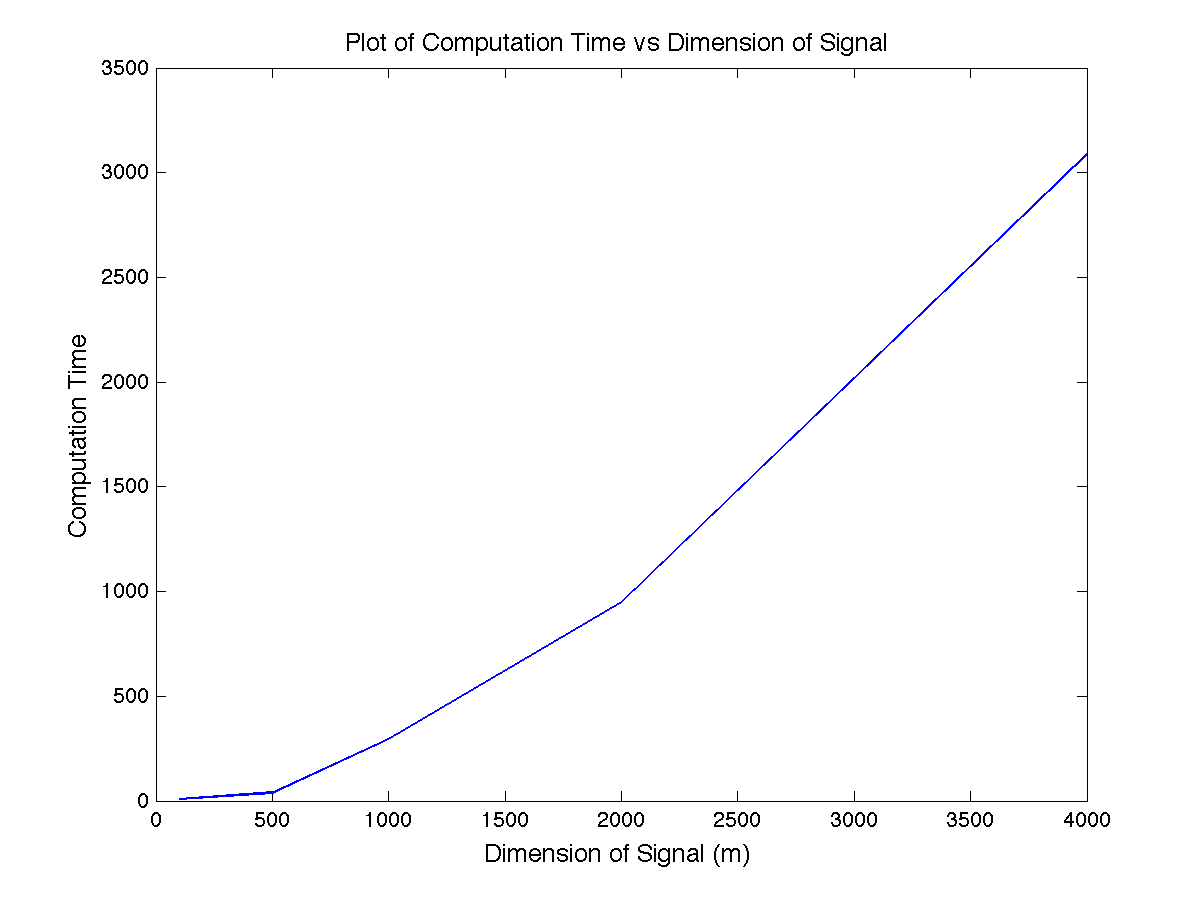}
\includegraphics[width=2.3in]{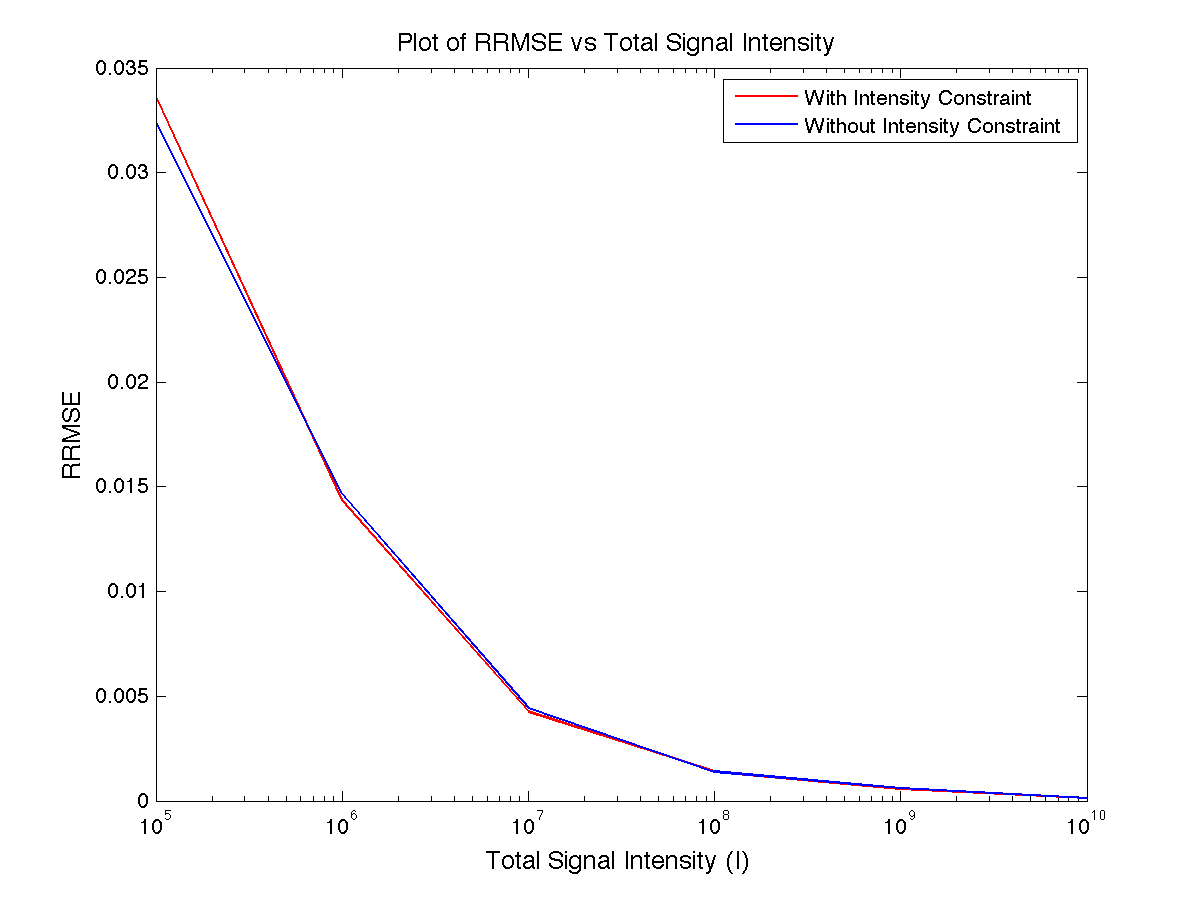}
\caption{Left: Time taken for the CVX solver on problem (P2) versus signal dimension $m$, Right: RMSE comparison for problem (P2) with and without imposition of the $\|\boldsymbol{x^\star}\|_1 = I$ constraint.}
\label{fig:time_complexity}
\end{figure*}

We tested the performance of (\textsf{P4}) on an image reconstruction task from compressed measurements under Poisson noise. Each patch of size $7 \times 7$ from a gray-scale image was vectorized and 25 Poisson-corrupted measurements were generated using the sensing matrix discussed in Section \ref{sec:Main_result}. This model is reminiscent of the architecture of the compressive camera designed in \cite{Oike2013} except that we considered overlapping patches here. Each patch was reconstructed from its compressed measurements independently by solving (\textsf{P4}) with sparsity in a 2D-DCT basis. The final image was reconstructed by averaging the reconstructions of overlapping patches. This experiment was repeated for different $I$ values by suitably rescaling the intensities of the original image. In Figure \ref{fig:results_image}, we show reconstruction results with (\textsf{P4}) under different values of $I$. There is a sharp decrease in relative reconstruction error with increase in $I$.
Note that in our experiments, we have not made use of the hard constraint $\|\boldsymbol{x^{\star}}\|_1 = I$  in problem (\textsf{P2}). In practice, we however observed that the estimated $\|\boldsymbol{x^\star}\|_1$ was close to the true $I$, especially for higher values of $I \geq 10^6$, and moreover even imposition of the constraint did not significantly alter the results as can be seen in Figure \ref{fig:time_complexity} for a 100-dimensional signal with 50 measurements and sparsity 5. Strictly speaking, the function $J(\boldsymbol{y},\boldsymbol{Ax})$ is not H\"older continuous due to the presence of entropy-like terms $y \log y$ that are undefined for $y = 0$, which affects the theoretical convergence guarantees for convex optimization. This issue can be alleviated by replacing $J(\boldsymbol{y},\boldsymbol{Ax})$ with $J(\boldsymbol{y}+\beta,\boldsymbol{Ax}+\beta)$ for some $\beta \approx 0, \beta > 0$, similar to \cite{Harmany2012} for the Poisson log-likelihood. In practice however, we set $\beta = 0$ and ignored all zero-valued measurements. This weeding out had to be performed very rarely for moderate or high $I$. Also, to get an idea of the computational complexity of the method, we plot a graph (Figure \ref{fig:time_complexity}) of the reconstruction time (till convergence) for signals of fixed sparsity 10 and dimensions $m$ ranging from 100 to 4000, with $N = m/2$ measurements in each case. 
Summarily, these numerical experiments confirm the efficacy of using the JSD/SQJSD in Poisson compressed sensing problems. In particular, the statistical properties of the SQJSD allow for compressive reconstruction with statistically motivated parameter selection, unlike methods based on the Poisson negative log-likelihood which require tweaking of the regularization/signal sparsity parameter.

\section{Relation to Prior Work}
\label{sec:prior_work}
There excellent algorithms for Poisson reconstruction such as \cite{Lingenfelter2009, Starck2010, Zhang2008, Sra2008}, but these methods do not provide performance bounds. In this section, we put our work in the context of existing work on Poisson compressed sensing with theoretical performance bounds. These techniques are based on one of the following categories: (a) optimizing either the Poisson negative log-likelihood (NLL) along with a regularization term, or (b) the LASSO, or (c) using the variance stabilization transform (VST). 
\subsection{Comparison with Poisson NLL based methods}
These methods include \cite{Raginsky2010,Jiang2015,Raginsky2011,Li2016_arxiv,Rohban2016,YHLi2015,Wang2015}. One primary advantage of the SQJSD-based approach over the Poisson NLL is that the former (unlike the latter) is a metric, \emph{and} can be bounded by values independent of $I$ as demonstrated in Section \ref{subsec:JSD_SQJSD}. In principle, this allows for an estimator that in practice does not require tweaking a regularization or signal sparsity parameter, and instead requires a statistically motivated bound $\epsilon$ to be specified, which is more intuitive. Moreover, the methods in \cite{Raginsky2010,Jiang2015} (and their extensions to the matrix completion problem in \cite{YXie2013,Cao2016,Soni2016}) employ $\ell_0$-regularizers for the signal, due to which the derived bounds are applicable only to computationally intractable estimators. The results in both papers have been presented using estimators with $\ell_1$ regularizers with the regularization parameters (as in \cite{Raginsky2010}) or signal sparisty parameter (as in \cite{Jiang2015}) chosen omnisciently, but the derived bounds are not applicable for the implemented estimator.  In contrast, our approach proves error bounds with the $\ell_1$ sparsity regularizer for which efficient and tractable algorithms exist. Moreover, the analysis in \cite{Jiang2015} is applicable to exactly sparse signals, whereas our work is applicable to signals that are sparse or compressible in any orthonormal basis. However the work in \cite{Jiang2015} does perform a lower bounds analysis, which we have not presented here. Recently, NLL-based tractable minimax estimators have been presented in \cite{Li2016_arxiv,Rohban2016}, but in both cases, knowledge of an upper bound on the signal sparsity parameter ($\ell_q$ norm of the signal, $0 < q \leq 1$) is required for the analysis, even if the sensing matrix were to obey the RIP. A technique for deriving a regularization parameter to ensure statistical consistency of the $\ell_1$-penalized NLL estimator has been proposed in \cite{YHLi2015}, but that again requires knowledge of the signal sparsity parameter.  In our work, the constraint $\|\boldsymbol{x}\|_1=I$ was required only due to the specific structure of the sensing matrix, and even there, it was not found to be necessary in practical implementation. For clarity the specific objective functions used in these techniques is summarized in Table \ref{tab:comparisons}. The work in \cite{Raginsky2011} deals with a \emph{specific} type of sensing matrices called the expander-based matrices, unlike the work in this paper which deals with any randomly generated matrices of the form Eqn. \ref{eq:Phi}, and the bounds derived in \cite{Raginsky2011} are only for signals that are sparse in the \emph{canonical} basis. In \cite{Wang2015}, performance bounds are derived \textit{in situ} with system calibration error estimates for \emph{multiple} measurements, which is essentially a different computational problem, which again requires knowledge of regularization parameters.
\begin{table}
\begin{center}
  \begin{tabular}{| c | p{13cm} |}
    \hline
	Method & Objective Function \\ \hline \hline
	This paper & Problem (P2) from Section \ref{subsec:main_c}, with $\epsilon$ chosen using properties of the SQJSD \\ \hline
	\cite{Raginsky2010} & $\textrm{NLL}(\*y,\*\Phi\*x) + \lambda \textrm{pen}(\*D^T\*x)$ such that $\* x \succeq \*0, \|\*x\|_1=I$ where $\textrm{pen}(\*D^T\*x) = \|\*D^T\*x\|_0$ \\ \hline
	\cite{Jiang2015} & $\textrm{NLL}(\*y,\*\Phi\*x)$ such that $\* x \succeq \*0, \|\*x\|_1=I, \|\*D^T\*x\|_0 \leq s$ for sparsity basis $\*D$ \\ \hline
	\cite{Rohban2016} & $\textrm{NLL}(\*y,\*\Phi\*x)$ such that $\* x \succeq \*0, \|\*D^T\*x\|_1 \leq s$ for sparsity basis $\*D$\\ \hline
	\cite{Li2016_arxiv} & $\textrm{NLL}(\*y,\*\Phi\*x)$ such that $\* x \succeq \*0, \|\*x\|_1=I, \|\*D^T\*x\|^q_q \leq s$ for sparsity basis $\*D$\\ \hline
	\cite{Garg2017_arxiv} & $\|\*D^T\*x\|_1$ such that $\|\sqrt{\*y}-\sqrt{\*\Phi\*x}\|_2 \leq \epsilon, \*x \succeq \*0, \|\*x\|_1=I$ for sparsity basis $\*D$ with $\epsilon$ picked based on chi-square tail bounds\\ \hline
	\cite{Jiang2015_arxiv} & $\|\*y-\*\Phi\*x\|^2 + \lambda \sum_{k} d_k (\*D^T \*x)_k$ for sparsity basis $\*D$, with weights $d_k$ picked statistically \\ \hline
	\cite{Rish2009} & $\|\*D^T\*x\|_1$ such that $\textrm{NLL}(\*y,\*\Phi \*x) \leq \epsilon$ where no criterion to choose $\epsilon$ is analyzed \\ \hline
	  \end{tabular}
\label{tab:comparisons}
\caption{Objective functions optimized by various Poisson compressed sensing methods}
\end{center}
\end{table}

\subsection {Comparison with LASSO-based methods}
These methods include \cite{Ivanoff2016,Jiang2015_arxiv,Blazere2014,Jia2013,Kakade2010,Rish2009}. The performance of the LASSO (designed initially for homoscedastic noise) under heterscedasticity associated with the Poisson noise model is examined in \cite{Jia2013} and necessary and sufficient conditions are derived for the sign consistency of the LASSO. Weighted/adaptive LASSO and group LASSO schemes with provable guarantees based on Poisson concentration inequalities have been proposed in \cite{Ivanoff2016,Jiang2015_arxiv}. Group LASSO based bounds have also been derived in \cite{Blazere2014} and applied to Poisson regression. Bounds on recovery error using an $\ell_1$ penalty are derived in \cite{Rish2009} and \cite{Kakade2010} based on the RIP and maximum eigenvalue condition respectively. These techniques do not provide bounds for realistic physical constraints in the form of flux-preserving sensing matrices. The quantity $\epsilon$ is not analyzed theoretically in \cite{Rish2009} unlike in our method - see Table \ref{tab:comparisons}. Moreover the LASSO is not a probabilistically motivated (i.e. penalized likelihood based) estimator for the case of Poisson noise. Even considering an approximation of $\textrm{Poisson}(\lambda)$ by $\mathcal{N}(\lambda,\lambda)$, the approximated likelihood function would be $K \triangleq \sum_{i=1}^n \frac{(y_i-[\boldsymbol{Ax}]_i)^2}{[\boldsymbol{Ax}]_i} + \log [\boldsymbol{Ax}]_i$ (which is non-convex in $\boldsymbol{x}$) and not $\sum_{i=1}^n (y_i-[\boldsymbol{Ax}]_i)^2$ as considered in the LASSO. However $J(\*y,\*A\*x)$ is a convex function, which is a lower bound on $K$ if $[\boldsymbol{Ax}]_i \geq 1$ as shown in Eqn. \ref{step3_1} while proving Theorem 1. Therefore our SQJSD method provides a tractable way to implement such a non-convex variant of the LASSO under some mild restrictions on the measurements. 

\subsection{Comparison with VST-based methods}
VST-based methods, especially those based on variants of the square-root transformations, have been used extensively in denoising \cite{Makitalo2013} and deblurring \cite{Dupe2009} but without performance bounds. In the context of Poisson CS, the VST converts a linear problem into a non-linear one. However, our group has recently shown the advantages of the VST for Poisson CS reconstructions in \cite{Garg2017, Garg2017_arxiv} with similar statistically motivated parameter selection. However in this paper, we present the result that the SQJSD also possesses such variance stabilizing properties for the Poisson distribution.

\section{Relation between the JSD and a Symmetrized Poisson Negative Log Likelihood}
\label{sec:relation}
In this section, we demonstrate the relationship between the JSD and an approximate symmetrized version of the Poisson negative log likelihood function. Consider an underlying noise-free signal $\boldsymbol{x} \in {\mathbb{R}_{+}}^{m \times 1}$. Consider that a compressive sensing device acquires $N \ll m$ measurements of the original signal $\boldsymbol{x}$ to produce a measurement vector $\boldsymbol{y} \in {\mathbb{Z_{+}}}^{N \times 1}$. Assuming independent Poisson noise in each entry of $\boldsymbol{y}$, we have $\forall i, 1 \leq i \leq N, {y}_i \sim \textrm{Poisson}(\boldsymbol{\Phi x})_i$, where as considered before, $\boldsymbol{\Phi}$ is a non-negative flux-preserving sensing matrix. The main task is to estimate the original signal $\boldsymbol{x}$ from $\boldsymbol{y}$. A common method is to maximize the following likelihood in order to infer $\boldsymbol{x}$:
\begin{equation}
\begin{split}
\mathcal{L}(\boldsymbol{y}|\boldsymbol{\Phi x}) &= \prod_{i=1}^{N} p({y}_i|(\boldsymbol{\Phi x})_i) \\ &= \prod_{i=1}^{N} \dfrac{{(\boldsymbol{\Phi x})_i}^{{y}_{i}}}{{{y}_{i}}!} e^{{-(\boldsymbol{\Phi x})_i}}.
\end{split}
\end{equation}
The negative log-likelihood $\mathcal{NLL}$ can be approximated as:
\begin{equation} \mathcal{NLL}(\boldsymbol{y},\boldsymbol{\Phi x}) \approx \sum_{i=1}^N y_i \log{\dfrac{y_i}{(\boldsymbol{\Phi x})_i} - y_i + (\boldsymbol{\Phi x})_i} + \dfrac{\log y_i}{2} + \dfrac{\log 2\pi}{2}.\end{equation}
This expression stems from the Stirling's approximation \cite{StirlingWiki} for $\log y_i!$ given by
\begin{equation}
\log y_i! \approx y_i \log y_i - y_i + \dfrac{\log y_i}{2} + \dfrac{\log 2\pi}{2}.
\end{equation}
This is derived from Stirling's series given below as follows for some integer $n \geq 1$:
\begin{equation}
n! \approx \sqrt{2\pi n}\big(\dfrac{n}{e}\big)^n \big(1 + \dfrac{1}{12n} + \dfrac{1}{288n^2}\big) \approx \sqrt{2\pi n}\big(\dfrac{n}{e}\big)^n.
\end{equation}
Consider the generalized Kullback-Leibler divergence between $\boldsymbol{y}$ and $\boldsymbol{\Phi x}$, denoted as $G(\boldsymbol{y}, \boldsymbol{\Phi x})$ and defined as
\begin{equation} G(\boldsymbol{y}, \boldsymbol{\Phi x}) \triangleq \sum_{i=1}^N y_i \log{\dfrac{y_i}{(\boldsymbol{\Phi x})_i} - y_i + (\boldsymbol{\Phi x})_i} .\label{eq:G} \end{equation}
The generalized Kullback-Leibler divergence turns out to be the Bregman divergence for the Poisson noise model \cite{Collins2001} and is used in maximum likelihood fitting and non-negative matrix factorization under the Poisson noise model \cite{Fevotte2009}.
The negative log-likelihood can be expressed in terms of the generalized Kullback-Leibler divergence in the following manner:
\begin{equation}
\mathcal{NLL}(\boldsymbol{y},\boldsymbol{\Phi x}) \approx G(\boldsymbol{y},\boldsymbol{\Phi x}) + \sum_{i=1}^N \Big(\dfrac{\log y_i}{2}+\dfrac{\log 2\pi}{2}\Big).
\end{equation}
Let us consider the following symmetrized version of the $\mathcal{NLL}$:
\begin{eqnarray}
\mathcal{SNLL}(\boldsymbol{y},\boldsymbol{\Phi x}) = \mathcal{NLL}(\boldsymbol{y},\boldsymbol{\Phi x}) +\mathcal{NLL}(\boldsymbol{\Phi x},\boldsymbol{y}) \approx G(\boldsymbol{y},\boldsymbol{\Phi x}) + G(\boldsymbol{\Phi x},\boldsymbol{y}) 
+ \sum_{i=1}^N \Big(\dfrac{\log y_i}{2} +\dfrac{\log (\boldsymbol{\Phi x})_i}{2} + \log 2\pi \Big) \\ \nonumber
\geq G(\boldsymbol{y},\boldsymbol{\Phi x}) + G(\boldsymbol{\Phi x},\boldsymbol{y}) = D(\boldsymbol{y},\boldsymbol{\Phi x}) + D(\boldsymbol{\Phi x},\boldsymbol{y}),
\label{eq:NLL_s_D}
\end{eqnarray}
where $D$ is the Kullback-Leibler divergence from Eqn. \ref{eq:D}. The inequality above is true when the term in parantheses is non-negative, which is true when either (1) for each $i$, we must have $y_i \geq \dfrac{1}{4 \pi^2 (\boldsymbol{\Phi x})_i}$, or (2) the \emph{minimum} value for $y_i \geq d \triangleq \dfrac{1}{4\pi^2 \big(\prod_{i=1}^N (\boldsymbol{\Phi x})_i\big)^{(1/N)}}$. We collectively denote these conditions as `Condition 1' henceforth. Note that, given the manner in which $\boldsymbol{\Phi}$ is constructed, we have the guarantee that $(\boldsymbol{\Phi x})_i \geq \dfrac{x_{min}}{N}$ with a probability of $1 - Np^m$ where $x_{min}$ is the minimum value in $\boldsymbol{x}$. The quantity on the right hand side of the last equality above follows from Eqns. \ref{eq:D} and \ref{eq:G}, and yields a symmetrized form of the Kullback-Leibler divergence $D_s(\boldsymbol{y},\boldsymbol{\Phi x})\triangleq D(\boldsymbol{y},\boldsymbol{\Phi x}) + D(\boldsymbol{\Phi x},\boldsymbol{y})$. Now, we have the following useful lemma giving an inequality relationship between $D_s$ and $J$, the proof of which follows \cite{Lin1991} and can be found in the supplemental material in Section \ref{sec:suppmat}. \\
\textbf{Lemma 3:} Given non-negative vectors $\boldsymbol{u}$ and $\boldsymbol{v}$, we have $\dfrac{1}{4}D_s(\boldsymbol{u},\boldsymbol{v}) \geq J(\boldsymbol{u},\boldsymbol{v})$.

Combining Eqns. \ref{eq:NLL_s_D} and Lemma 3, we arrive at the following conclusion if `Condition 1' holds true:
\begin{eqnarray}\mathcal{SNLL}(\boldsymbol{y},\boldsymbol{\Phi x}) \leq \epsilon \implies J(\boldsymbol{y},\boldsymbol{\Phi x}) \leq \epsilon/4 
\implies \sqrt{J(\boldsymbol{y},\boldsymbol{\Phi x})} \leq \epsilon' \triangleq \sqrt{\epsilon}/2.
\label{eq:SNLL_SQJSD}
\end{eqnarray}
 
Let us consider the following optimization problem:
\begin{eqnarray}
\textrm{(\textsf{P3}): minimize} \|\boldsymbol{z}\|_1 \textrm{ such that } \mathcal{SNLL}(\boldsymbol{y},\boldsymbol{A z}) \leq \epsilon, \boldsymbol{z} \succeq \boldsymbol{0}, \|\boldsymbol{\Psi z}\|_1 = I.
\label{eq:l1_SNLL}
\end{eqnarray}
Following Eqn. \ref{eq:SNLL_SQJSD}, we observe that a solution to (\textsf{P3}) is also a solution to (\textsf{P2}) with slight abuse of notation (\textit{i.e.}, the $\epsilon$ in (\textsf{P2}) should actually be $\epsilon'$ defined in Eqn. \ref{eq:SNLL_SQJSD}). Note that Condition 1 can fail with higher probability if $(\boldsymbol{\Phi x})_i$ is small, due to which the $J \leq \mathcal{SNLL}$ bound may no longer hold. However, this does not affect the validity of Theorem 1.3 or the properties of the estimator proposed in this paper. Note that we choose to solve (\textsf{P2}) instead of (\textsf{P3}) in this paper, as the SQJSD and not $\mathcal{SNLL}$ is a metric, which makes it easier to establish theoretical bounds using SQJSD. 

\section{Conclusion}
\label{sec:conclusion}
In this paper, we have presented new upper bounds on the reconstruction error from compressed measurements under Poisson noise in a realistic imaging system obeying the non-negativity and flux-preservation constraints, for a \emph{computationally tractable} estimator using the $\ell_1$ norm sparsity regularizer. Our bounds are easy to derive and follow the skeleton of the technique laid out in \cite{Candes2008}. The bounds are based interesting properties of the SQJSD from Section \ref{subsec:JSD_SQJSD}, some of which are derived in this paper, and are applicable to sparse as well as compressible signals in any chosen orthonormal basis. We have presented numerical simulations with parameters chosen based on noise statistics (unlike the choice of regularization or signal sparsity parameters in other techniques), showing the efficacy of the method in reconstruction from compressed measurements under Poisson noise.  We observe that the derived upper bounds decrease with an increase in the original signal flux, i.e. $I$. However the bounds do not decrease with an increase in the number of measurements $N$, unlike conventional compressed sensing. This observation, though derived independently and using different techniques, agrees with existing literature on Poisson compressed sensing or matrix completion \cite{Raginsky2010,Cao2016,YXie2013}. The reason for this strange observation is the division of the signal flux across the $N$ measurements, thereby leading to poorer signal to noise ratio per measurement.

There exist several avenues for future work, as follows. A major issue is to explore theoretical error bounds in the absence of the knowledge of $I$, which is an open problem in flux-preserving systems to the best of our knowledge (even though we have excellent numerical results without knowing $I$). Furthermore, it will be useful to derive lower-bounds on the reconstruction error and extend our theory to the problem of matrix completion under Poisson noise.

\section{Appendix}
\label{sec:proofs}
\subsection{Proof of Theorem 1}
To prove this theorem, we first begin by considering $y \sim \textrm{Poisson}(\gamma)$ where $\gamma \in \mathbb{R}$ and derive bounds for the mean and variance of $J(y,\gamma)$. Thereafter, we generalize to the case with multiple measurements. 
\\
Let $f(y) \triangleq J(y,\gamma)$. Hence we have
 \begin{align*} 
f(y) &=\frac{1}{2}(\gamma\log{\gamma}+y\log{y}) - \frac{\gamma+y}{2} \log \Big(\frac{\gamma+y}{2}\Big). \\
\therefore f^{(1)}(y) &= \frac{1}{2}[\log{y} - \log \Big(\frac{\gamma+y}{2}\Big)]. \\
\therefore  f(y) &= \int_{x}^{y} f^{(1)}(t) dt  \text{ as  } f(\gamma) = 0. 
\end{align*}
where $f^{(k)}(y)$ stands for the $k^{\textrm{th}}$ derivative of $f(y)$. As $f^{(1)}(y)$ is a non-decreasing function (since $f^{(2)}(y)$ is non-negative for all $y$), we have 
\begin{equation}  f(y) \leq (y-\gamma) f^{(1)}(y). \label{step3} \end{equation}
Likewise, noting that $f^{(1)}(\gamma)=0$ we get $f^{(1)}(y) = \int_{\gamma}^{y} f^{(2)}(t) dt$. We know that $f^{(2)}(y) = \frac{1}{2}\Big[ \frac{1}{y} - \frac{1}{y + \gamma}\Big]$ is a decreasing function as $f^{(3)}(y)$ is negative for all $y$.
\\
If $y \geq \gamma$ then $f^{(2)}(y) \leq f^{(2)}(\gamma)$. Therefore, $f^{(1)}(y) \leq (y-\gamma)f^{(2)}(\gamma)$. If $y \leq \gamma$ then $f^{(2)}(y) \geq f^{(2)}(\gamma)$. Therefore, $-f^{(1)}(y) \geq (\gamma-y)f^{(2)}(\gamma)$.
Combining Eqn. \ref{step3} with the above inequality, we get
 \begin{equation}  f(y) \leq (y-\gamma)^2 f^{(2)}(\gamma) = \frac{1}{4\gamma} (y-\gamma)^2. \label{step3_1} \end{equation}
Therefore, using $E[(y-\gamma)^2] = \gamma$ for a Poisson random variable, we have
 \begin{equation}  E[f(y)] \leq \frac{1}{4\gamma} E[(y-\gamma)^2] = \frac{1}{4}. \label{eq:eval_f} \end{equation}
Thus, we have found an upper bound on $E[f(y)]$ which is independent of $\gamma$. 
\\
We will now derive a lower bound on $E[f(y)]$, as it will be useful in deriving an upper bound for $\textrm{Var}(f(y))$. We can expand $f(y)$ using a second order Taylor series about $\gamma$ along with a (third order) Lagrange remainder term as follows:
 \begin{align*}  f(y) &= f(\gamma) + f^{(1)}(\gamma)(y-\gamma) + \frac{f^{(2)}(\gamma)}{2!} (y-\gamma)^2 +  \frac{f^{(2)}(z(y))}{3!} (y-\gamma)^3 \\ &=\frac{1}{8\gamma} (y-\gamma)^2 - \frac{1}{12} (y-\gamma)^3 \Big[ \frac{1}{z^2(y)} - \frac{1}{(\gamma+z(y))^2} \Big] \end{align*} 
for some $z(y)$ that lies in the interval $(y, \gamma)$ or $(\gamma, y)$. 
Therefore, 
 \begin{align*} E[f(y)] &= \frac{1}{8\gamma} E[(y-\gamma)^2] - \frac{1}{12} \Big[ \sum^{\infty}_{y=0} \frac{e^{-\gamma}{\gamma}^{y}}{y!} (y-\gamma)^3 \Big[ \frac{1}{z^2(y)} - \frac{1}{(\gamma+z(y))^2} \Big] \Big] \\
 &=  \frac{1}{8} - \frac{1}{12} \Big[ \sum^{\infty}_{y=0} \frac{e^{-\gamma}{\gamma}^{y}}{y!} (y-\gamma)^3 \Big( \frac{1}{z^2(y)} - \frac{1}{(\gamma+z(y))^2} \Big) \Big]. 
 \end{align*}

Let $\alpha$ be the largest integer less than or equal to $\gamma$. We can split the second term in the RHS of the above expression into the sum of two terms $I_1$ and $-I_2$, depending upon whether $y$ is greater than $\alpha$ or not. $I_1$ and $I_2$ are defined as follows:
%\begin{\align*} I_1 = \frac{1}{12} \Big[ \sum^{\infty}_{\*y=0} \frac{e^{-\*x}{\*x}^{\*y}}{\*y!} (\*y-\*x)^3 \Big( \frac{1}{\*z^2} - \frac{1}{(\*x+\*z)^2} \Big) \Big] \end{\align*}
\begin{align*} I_1 &= \frac{1}{12} \Big[ \sum^{\alpha}_{y=0} \frac{e^{-\gamma}{\gamma}^{y}}{y!} (\gamma-y)^3 \Big( \frac{1}{z^2(y)} - \frac{1}{(\gamma+z(y))^2} \Big) \Big]  \\
I_2 &= \frac{1}{12} \Big[ \sum^{\infty}_{y=\alpha + 1} \frac{e^{-\gamma}{\gamma}^{y}}{y!} (y-\gamma)^3 \Big( \frac{1}{z^2(y)} - \frac{1}{(\gamma+z(y))^2} \Big) \Big]. 
\end{align*}

In order to lower bound $E[f(y)]$, we want to minimize $I_1$ and maximize $I_2$ w.r.t. $z(y)$. Since $\frac{1}{z^2(y)} - \frac{1}{(\gamma+z(y))^2}$ is a decreasing function of $z(y)$, it can be proved that $I_1$ is minimized when $z(y) = \gamma$ and that $I_2$ attains a maximum when $z(y) = \gamma$. Therefore, we obtain
\begin{equation} E[f(y)]  \geq \frac{1}{8} - \frac{1}{16\gamma^2}E[(y-\gamma)^3] = \frac{1}{8} - \frac{1}{16\gamma}. \label{step5} \end{equation}
This lower bound is loose if $\gamma < 0.5$ since we know that $E[f(y)]$ must clearly be non-negative. Hence it is more apt to express the lower bound as follows:
\begin{equation} E[f(y)]  \geq \textrm{max}(0,\frac{1}{8} - \frac{1}{16\gamma}). \label{step5a} \end{equation}
In summary, we have
\begin{equation} \textrm{max}(0,\frac{1}{8} - \frac{1}{16\gamma}) \leq E[f(y)] \leq \frac{1}{4}.  \end{equation}
We now proceed to derive an upper bound on the variance of $f(y)$. 
\\
Using Eqn. \ref{step3_1} we get,
\begin{equation*} E[(f(y))^2] \leq \frac{1}{16\gamma^2} E[(y-\gamma)^4]  = \frac{\gamma(1+3\gamma)}{16\gamma^2} \leq  \frac{3}{16} + \frac{1}{16\gamma}. \label{step6_a} \end{equation*}
Recall that $\mathrm{Var}[f(y)]=E[(f(y))^2] - (E[f(y)])^2$. Using Eqn. \ref{step5} and \ref{step6_a}, we get
\begin{eqnarray} 
\mathrm{Var}(f(y)) &\leq \frac{3}{16} + \frac{1}{16\gamma}  - \Big(\textrm{max}(0,\Big[ \frac{1}{8} - \frac{1}{16\gamma} \Big])\Big)^2 \\
 &\leq \textrm{max}(0,\frac{11}{64} + \frac{5}{64\gamma} - \frac{1}{256\gamma^2}) \\
& \leq \frac{11}{64} + \frac{5}{64\gamma}. \label{step7}
\end{eqnarray}
Now consider that $\boldsymbol{y}$ is a vector of $N$ measurements such that $\forall i \in \{1,2,...,N\}, y_i \sim \textrm{Poisson}(\gamma_i)$ and all measurements are independent. We will later replace $\gamma_i$ by $(\boldsymbol{\Phi x})_i$ where $\boldsymbol{\Phi}$ is a non-negative flux-preserving matrix and $\*x$ is the unknown signal to be estimated. Let us define some terminology as follows:
\begin{align*}   
f_i(y_i) \triangleq \frac{(\gamma_i\log{\gamma_i}+y_i\log{y_i})}{2} - \frac{\gamma_i+y_i}{2} \log \Big(\frac{\gamma_i+y_i}{2}\Big), f(\boldsymbol{y}) \triangleq \sum^{N}_{i=1} f_i(y_i), g(\*y) \triangleq \sqrt{f(\*y)}.  
\end{align*}
Jensen's inequality gives the following upper bound on the expected value of $g(\*y)$:
\begin{equation} E[g(\*y)] = E[\sqrt{f(\*y)}] \leq \sqrt{ \sum^{N}_{i=1} E[f_i(y_i)]} \leq \sqrt{\frac{N}{4}}.\end{equation}
In order to lower bound $E[g(\*y)]$ we use the following inequality for the non-negative variable $f$:
\begin{equation*} \sqrt{f} \geq 1 + \frac{f-1}{2} - \frac{(f-1)^2}{2}. \end{equation*}
This inequality follows since it is equivalent to $3f-f^2\leq 2\sqrt{f}$ which implies $3b-b^3 \leq 2$ which is true for any $b \geq 0$.
Define $\tilde{f} \triangleq \dfrac{f}{E[f]}$ such that $E[\tilde{f}]=1$. Therefore, we have the following inequalities:
\begin{align*} 
\sqrt{\tilde{f}} &\geq 1 + \frac{\tilde{f}-1}{2} - \frac{(\tilde{f}-1)^2}{2} \\
\therefore E[\sqrt{\tilde{f}}] &\geq 1 - \frac{\mathrm{Var}(\tilde{f})}{2} \\
\therefore E[\sqrt{f}] &\geq  \sqrt{E[f]} \Big(1 - \frac{\mathrm{Var}(f)}{2E[f]^2}  \Big)\\
\therefore E[g] &\geq  \sqrt{E[f]} \Big(1 - \frac{\mathrm{Var}(f)}{2E[f]^2}  \Big).
\end{align*}

Now, we can find an upper bound on $\mathrm{Var}[g(\*y)]$
 \begin{align*} 
 \mathrm{Var}(g) &= E[g^2] - E[g]^2 \\
 &\leq E[f] - E[f]\Big(1 - \frac{\mathrm{Var}(f)}{2E[f]^2}  \Big)^2 \\
 &\leq \frac{\mathrm{Var}(f)}{E[f]} - \frac{1}{4} \frac{\mathrm{Var}(f)^2}{E[f]^3}.
\end{align*}

As for different $i$, the variables $f_i(y_i)$ are independent of each other, we get $\mathrm{Var}(f) =  \sum^{N}_{i=1} \mathrm{Var}(f_i)$, due to which we have:
\begin{align*}   \
\mathrm{Var}(g) & \leq \frac{ \sum^{N}_{i=1} \mathrm{Var}(f_i) }{ \sum^{N}_{i=1} E(f_i) } - \frac{1}{4} \frac{(\sum^{N}_{i=1} \mathrm{Var}(f_i))^2}{(\sum^{N}_{i=1} E(f_i))^2} \\
&\leq \dfrac{11N+5 \sum_{i=1}^N 1/\gamma_i}{\textrm{max}(0,4(2N-\sum_{i=1}^N 1/\gamma_i))}.
\end{align*}
The last step follows from Eqn. \ref{step7} and \ref{step5a}.  
Now we consider replacing $\gamma_i$ by $(\boldsymbol{\Phi x})_i$. Since $\boldsymbol{\Phi}$ contains the values 0 or $\frac{1}{N}$, we see that $s_i = N \times (\boldsymbol{\Phi x})_i$ is the summation of a subset of the elements in the vector $\boldsymbol{x}$. This gives us the final upper bound
\begin{equation}
\textrm{Var}[\sqrt{J(\* y, \*\Phi\*x)}] \leq \dfrac{11+5 \sum_{i=1}^N 1/s_i}{\textrm{max}(0,4(2-\sum_{i=1}^N 1/s_i))}.
\end{equation}
In order to obtain a tail bound on $\sqrt{J(\* y, \*\Phi\*x)}$, we can use Chebyshev's inequality to prove that $P(\sqrt{J(\* y, \*\Phi\*x)} \leq \sqrt{N/4} + \sigma\sqrt{N}) \geq 1-\frac{1}{N}$, where $\sigma^2$ is the variance of $f_i$ and is upper bounded by (approximately) $\frac{11}{64}$. However, we show here that $\sqrt{J(\* y, \*\Phi\*x)}$ is approximately Gaussian distributed which leads to an even higher probability. By the central limit theorem, we know that $P(\frac{f(\*y)-N\mu}{\sigma\sqrt{N}} \leq \alpha) \rightarrow \Phi_g(\alpha)$ as $N \rightarrow \infty$, where $\Phi_g$ is the CDF for $\mathcal{N}(0,1)$, and $\mu$ is the expected value of $f_i$. All the $f_i$ values will have near-identical variances ($\leq 11/64$ from Eqn. \ref{step7}) if the intensity of the measurements is sufficiently high. Due to the continuity of $\Phi_g$, we have $P(\frac{f(\*y)-N\mu}{\sigma\sqrt{N}} \leq \alpha + \frac{\alpha^2 \sigma^2}{4\mu\sigma\sqrt{N}}) \rightarrow \Phi_g(\alpha)$ as $N \rightarrow \infty$. Hence we have $P(f(\*y) \leq (\sqrt{N\mu}+\frac{\alpha \sigma}{2\sqrt{\mu}})^2) \rightarrow \Phi_g(\alpha)$ as $N \rightarrow \infty$, and taking square roots we get $P(\sqrt{f(\*y)} \leq (\sqrt{N\mu}+\frac{\alpha \sigma}{2\sqrt{\mu}})) \rightarrow \Phi_g(\alpha)$ as $N \rightarrow \infty$. By rearrangement, we obtain $P(\frac{\sqrt{f(\*y)}-\sqrt{N\mu}}{\sigma/(2\sqrt{\mu})} \leq \alpha) \rightarrow \Phi_g(\alpha)$ as $N \rightarrow \infty$. With this development and since $\mu \leq 1/4$ from Eqn. \ref{eq:eval_f}, we can now invoke a Gaussian tail bound to establish that $P(\sqrt{J(\* y, \*\Phi\*x)} \leq \sqrt{N/4} + \sigma\sqrt{N}) \geq 1-2e^{-N/2}$. Note that the Gaussian nature of $\sqrt{J(\* y, \*\Phi\*x)}$ emerges from the central limit theorem and is only an asymptotic result. However we consistently observe it to be true even for small values of $N \sim 10$ as confirmed by a Kolmogorov-Smirnov test (see \cite{suppcode}). $\Box$

\subsection{Proof of Theorem 2}
Our proof follows the approach for the proof of the key results in \cite{Candes2008,Studer2012} for the case of bounded, signal-independent noise, but meticulously adapted here for the case of Poisson noise.
\begin{enumerate}
\item Consider an upper bound $\epsilon$ on $\sqrt{J(\boldsymbol{y},\boldsymbol{\Phi x})}$, \textit{i.e.}, $\sqrt{J(\boldsymbol{y},\boldsymbol{\Phi x})} \leq \epsilon$. We will later set $\epsilon$ using tail bounds on the distribution of the random variable $\sqrt{J(\boldsymbol{y},\boldsymbol{\Phi x})}$. For now, we prove the following result:
\begin{equation}\|\boldsymbol{\Phi\Psi(\theta-\theta^{\star})}\|_2 \leq 2\sqrt{8I}\epsilon \label{epsilon}.\end{equation}
We have 
\begin{align*}
\|\boldsymbol{\Phi\Psi\theta- \Phi\Psi\theta^{\star}}\|_2  \leq \|\boldsymbol{\Phi\Psi(\theta-\theta^{\star})}\|_1 = I\|\boldsymbol{\Phi\Psi}(\dfrac{\boldsymbol{\theta}}{I}-\dfrac{\boldsymbol{\theta^\star}}{I})\|_1 \\ \nonumber
\leq I\sqrt{8J(\dfrac{\boldsymbol{\Phi\Psi\theta}}{I}, \dfrac{\boldsymbol{\Phi\Psi\theta^\star}}{I})}  \quad \textrm{by Lemma 2} \\ \nonumber
= I\sqrt{8J(\dfrac{\boldsymbol{\Phi\Psi\theta}}{I}, \dfrac{\boldsymbol{y}}{I})} + I\sqrt{8J(\dfrac{\boldsymbol{\Phi\Psi\theta^\star}}{I}, \dfrac{\boldsymbol{y}}{I})} \textrm{ by Lemma 1 } \\ \nonumber
= \dfrac{I}{\sqrt{I}}\sqrt{8J(\boldsymbol{\Phi\Psi\theta}, \boldsymbol{y})} + \dfrac{I}{\sqrt{I}}\sqrt{8J(\boldsymbol{\Phi\Psi\theta^\star}, \boldsymbol{y})}  \leq 2\sqrt{8I}\epsilon. \label{eq:8e}\\ \nonumber
\end{align*}
Note that Lemma 2 can be used in the third step above because we have imposed the constraint that $\|\boldsymbol{\Psi \theta^{\star}}\|_1 = \|\boldsymbol{\Psi \theta}\|_1 = I$ and because by the flux-preserving property of $\boldsymbol{\Phi}$, we have $\|\boldsymbol{\Phi \Psi \theta}\|_1 \leq I$ and $\|\boldsymbol{\Phi \Psi \theta^{\star}}\|_1 \leq I$.

\item Let us define vector $\boldsymbol{h} \triangleq \boldsymbol{\theta^{\star}-\theta}$ which is the difference between the estimated and true coefficient vectors. Let us denote vector $\boldsymbol{h}_T$ as the vector equal to $\boldsymbol{h}$ only on an index set $T$ and zero at all other indices. Let $T^c$ denote the complement of the index set $T$. Let $T_0$ be the set of indices containing the $s$ largest entries of $\boldsymbol{h}$ (in terms of absolute value), $T_1$ be the set of indices of the next $s$ largest entries of {$\boldsymbol{h}_{T^c_0}$}, and so on. We will now decompose $\boldsymbol{h}$ as the sum of $\boldsymbol{h_{T_0}}, \boldsymbol{h}_{T_1}, \boldsymbol{h}_{T_2}, ...$. Our aim is to prove a logical and intuitive bound for both $\|\boldsymbol{h}_{T_0 \cup T_1}\|_2$ and $\|\boldsymbol{h}_{(T_0 \cup T_1)^c}\|_2$.
\item We will first prove the bound on $\|\boldsymbol{h}_{(T_0 \cup T_1)^c}\|_2$, in the following way:
\begin{enumerate}
\item  
We have 
\begin{align*} \|\boldsymbol{h}_{T_j}\|_2 = \sqrt{\sum_{k}\boldsymbol{h}_{{T_j}_k}^2} \leq s^{1/2} \|\boldsymbol{h}_{T_j}\|_\infty, \\  s\|\boldsymbol{h}_{T_j}\|_\infty \leq \sum_{i} |\boldsymbol{h}_{T_{{j-1}_i}}| = \|\boldsymbol{h}_{T_{j-1}}\|_1. \end{align*}
\\ Therefore, 
\begin{align*}\|\boldsymbol{h}_{T_j}\|_2 \leq s^{1/2} \|\boldsymbol{h}_{T_j}\|_\infty \leq s^{-1/2} \|\boldsymbol{h}_{T_{j-1}}\|_1. \end{align*}

\item 
Using Step 3(a), we get 
\begin{align*}
\|\boldsymbol{h}_{(T_0 \cup T_1)^c}\|_2 = \|\sum_{j \geq 2} \boldsymbol{h}_{T_j}\|_2 &\leq \sum_{j \geq 2} \|\boldsymbol{h}_{T_j}\|_2 
\\ &\leq s^{-1/2} \sum_{i\geq1} \|\boldsymbol{h}_{T_{i}}\|_1 \\&\leq s^{-1/2} \|\boldsymbol{h}_{(T_0)^c}\|_1.
\end{align*}

\item Using the reverse triangle inequality and the fact that $\boldsymbol{\theta}^{\star}$ is the solution of (\textsf{P2}), we have 
\begin{align*} \|\boldsymbol{\theta}\|_1 &\geq \|\boldsymbol{\theta}+\boldsymbol{h}\|_1 \\ & = \sum_{i \in T_0} |\theta_i + h_i| + \sum_{i \in {(T_0)}^c} |\theta_i + h_i| \\& \geq \|\boldsymbol{\theta}_{T_0}\|_1 - \|\boldsymbol{h}_{T_0}\|_1 + \|\boldsymbol{h}_{{(T_0)}^c}\|_1 - \|\boldsymbol{\theta}_{{(T_0)^c}}\|_1.\end{align*}
Rearranging the above equation gives us
\begin{align*}
\|\boldsymbol{h}_{{(T_0)}^c}\|_1 \leq  \|\boldsymbol{h}_{{(T_0)}}\|_1  + 2\|\boldsymbol{\theta}-\boldsymbol{\theta_s}\|_1
\end{align*}

\item We have 
\begin{align*}
 \|\boldsymbol{h}_{(T_0 \cup T_1)^c}\|_2 &\leq  s^{-1/2}\|\boldsymbol{h}_{(T_0)^c}\|_1 \\ &\leq  s^{-1/2}(\|\boldsymbol{h}_{{(T_0)}}\|_1  + 2\|\boldsymbol{\theta}-\boldsymbol{\theta_s}\|_1) \\ &\leq  \|\boldsymbol{h}_{{(T_0)}}\|_2  + 2s^{-1/2}\|\boldsymbol{\theta}-\boldsymbol{\theta_s}\|_1
\end{align*}
Using $\|\boldsymbol{h}_{{(T_0)}}\|_2  \leq \|\boldsymbol{h}_{T_0 \cup T_1}\|_2 $, we get 
\begin{align} \|\boldsymbol{h}_{(T_0 \cup T_1)^c}\|_2 \leq \|\boldsymbol{h}_{T_0 \cup T_1}\|_2 + 2s^{-1/2} \|\boldsymbol{\theta}-\boldsymbol{\theta_s}\|_1. 
 \end{align}
\end{enumerate}

\item We will now prove the bound on $\|\boldsymbol{h}_{(T_0 \cup T_1)}\|_2$, in the following way:
\begin{enumerate}
\item We have 
\begin{align*}
\boldsymbol{\Phi} &= \sqrt{\dfrac{p(1-p)}{N}} \boldsymbol{\tilde{\Phi}} + \dfrac{(1-p)}{N}\boldsymbol{1}_{N \times m} \\ \boldsymbol{\Phi\Psi(\theta-\theta^{\star})} &= \sqrt{\dfrac{p(1-p)}{N}} \boldsymbol{\tilde{\Phi}}\boldsymbol{\Psi}(\boldsymbol{\theta- \theta^{\star}}) + \\ & \qquad  \dfrac{(1-p)}{N}\boldsymbol{1}_{N \times m}\boldsymbol{\Psi}(\boldsymbol{\theta - \theta^{\star}}) \\
&=\sqrt{\dfrac{p(1-p)}{N}} \boldsymbol{\tilde{\Phi}}\boldsymbol{\Psi}(\boldsymbol{\theta- \theta^{\star}}) + \\ &\qquad
\dfrac{(1-p)}{N}(\|\boldsymbol{\Psi \theta}\|_1 - \|\boldsymbol{\Psi \theta^\star}\|_1) 
\end{align*}
As $\|\boldsymbol{\Psi\theta^{\star}}\|_1 = \|\boldsymbol{\Psi\theta}\|_1 = I$, we get
\begin{align} \boldsymbol{\Phi\Psi(\theta-\theta^{\star})} =  \sqrt{\dfrac{p(1-p)}{N}} \boldsymbol{\tilde{\Phi}}\boldsymbol{\Psi}(\boldsymbol{\theta- \theta^{\star}}).
 \label{eqB} \end{align}
   
Let us define $\boldsymbol{B}  \triangleq \boldsymbol{\tilde{\Phi}\Psi}$. If $N \geq O(s \log m)$, then $\boldsymbol{\tilde{\Phi}}$ obeys RIP of order $2s$ with very high probability, and so does the product $\boldsymbol{B}$ since $\boldsymbol{\Psi}$ is an orthonormal matrix \cite{Baraniuk2008}. 

From Eqn. \ref{eqB} above we have,
\begin{align*}  
 \|\boldsymbol{B(\theta-\theta^{\star})}\|_2 &= \sqrt{\dfrac{N}{p(1-p)}}  \|\boldsymbol{\Phi\Psi(\theta-\theta^{\star})}\|_2 \\
& \leq 2\sqrt{\dfrac{8NI}{p(1-p)}} \epsilon \textrm{ using Eqn. \ref{epsilon}} \\
\therefore  \|\boldsymbol{Bh}\|_2  &\leq  2\sqrt{\dfrac{8NI}{p(1-p)}} \epsilon 
\end{align*}

Defining $C_1 \triangleq 2\sqrt{\dfrac{8}{p(1-p)}}$, we have 
\begin{align}\|\boldsymbol{Bh}\|_2  \leq C_1\sqrt{NI}\epsilon \label{BhL2norm}\end{align}

\item The RIP of $\boldsymbol{B}$ with RIC $\delta_{2s}$ gives us, 
\begin{align*} \|\boldsymbol{Bh}_{T_0 \cup T_1}\|_2 \leq \sqrt{1+\delta_{2s}}\|\boldsymbol{h}_{T_0 \cup T_1}\|_2 \end{align*}
Using Eqn. \ref{BhL2norm} and the Cauchy-Schwartz inequality,
\begin{align} |\langle \boldsymbol{Bh}_{T_0 \cup T_1} , \boldsymbol{Bh}\rangle| &\leq \|\boldsymbol{Bh}_{T_0 \cup T_1}\|_2 \|\boldsymbol{Bh}\|_2 \nonumber \\ & \leq C_1\epsilon \sqrt{NI(1 + \delta_{2s})}  \|\boldsymbol{h}_{T_0 \cup T_1}\|_2.    \label{4b}\end{align}

\item Note that the vectors $\boldsymbol{h}_{T_0}$ and $\boldsymbol{h}_{T_j}$, $j \neq 0$ have disjoint support. Consider 
\begin{align*} |\langle \boldsymbol{Bh}_{T_0}, \boldsymbol{Bh}_{T_j}\rangle|= \|\boldsymbol{h}_{T_0}\|_2 \|\boldsymbol{h}_{T_j}\|_2|\langle \boldsymbol{B\hat{h}}_{T_0}, \boldsymbol{B\hat{h}}_{T_j}\rangle| \end{align*}
where $\boldsymbol{\hat{h}}_{T_0}$ and $\boldsymbol{\hat{h}}_{T_j}$ are unit-normalized vectors. 
This further yields,
\begin{align}  &|\langle \boldsymbol{Bh}_{T_0}, \boldsymbol{Bh}_{T_j}\rangle|   \nonumber \\
& =\|\boldsymbol{h}_{T_0}\|_2 \|\boldsymbol{h}_{T_j}\|_2 \dfrac{\|\boldsymbol{B}(\boldsymbol{\hat{h}}_{T_0}+\boldsymbol{\hat{h}}_{T_j})\|^2 - \|\boldsymbol{B}(\boldsymbol{\hat{h}}_{T_0}-\boldsymbol{\hat{h}}_{T_j})\|^2}{4} \nonumber \\
&\leq \|\boldsymbol{h}_{T_0}\|_2 \|\boldsymbol{h}_{T_j}\|_2  \dfrac{(1+\delta_{2s})(\|\boldsymbol{\hat{h}}_{T_0}\|^2+\|\boldsymbol{\hat{h}}_{T_j}\|^2)-(1-\delta_{2s})(\|\boldsymbol{\hat{h}}_{T_0}\|^2+\|\boldsymbol{\hat{h}}_{T_j}\|^2)}{4}  \nonumber \\  &\leq \delta_{2s} \|\boldsymbol{h}_{T_0}\|_2 \|\boldsymbol{h}_{T_j}\|_2.
\label{4c1}
\end{align}
Analogously, 
\begin{align} |\langle \boldsymbol{Bh}_{T_1}, \boldsymbol{Bh}_{T_j}\rangle| \leq \delta_{2s} \|\boldsymbol{h}_{T_1}\|_2 \|\boldsymbol{h}_{T_j}\|_2 \label{4c2}.\end{align}

\item We observe that 
\begin{align}\boldsymbol{Bh}_{T_0 \cup T_1} &= \boldsymbol{Bh} - \sum_{j \geq 2} \boldsymbol{Bh}_{T_j} \nonumber \\ \|\boldsymbol{Bh}_{T_0 \cup T_1}\|^2_2 &= \langle \boldsymbol{Bh}_{T_0 \cup T_1}, \boldsymbol{Bh}\rangle - \langle \boldsymbol{Bh}_{T_0 \cup T_1} , \sum_{j \geq 2} \boldsymbol{Bh}_{T_j}\rangle \label{4d}.\end{align}

\item Using the RIP of $\boldsymbol{B}$ and Eqns. \ref{4b}, \ref{4c1}, \ref{4c2}, \ref{4d}, we obtain
\begin{align*}
&(1-\delta_{2s})\|\boldsymbol{h}_{T_0 \cup T_1}\|^2_2  \leq \|\boldsymbol{Bh}_{T_0 \cup T_1}\|^2_2 \leq  C_1\epsilon \sqrt{NI(1 + \delta_{2s})}\|\boldsymbol{h}_{T_0 \cup T_1}\|_2 + \delta_{2s}(\|\boldsymbol{h}_{T_0}\|_2 + \|\boldsymbol{h}_{T_1}\|_2) \sum_{j \geq 2} \|\boldsymbol{h}_{T_j}\|_2.
\end{align*}
 As $\boldsymbol{h}_{T_0}$ and $\boldsymbol{h}_{T_1}$ are vectors with disjoint sets of non-zero indices, it follows that 
 \begin{align*} \|\boldsymbol{h}_{T_0}\|_2 + \|\boldsymbol{h}_{T_1}\|_2 \leq \sqrt{2} \|\boldsymbol{h}_{T_0 \cup T_1}\|_2. \end{align*}
Therefore, we get
\begin{align} &(1-\delta_{2s})\|\boldsymbol{h}_{T_0 \cup T_1}\|^2_2  \leq \|\boldsymbol{h}_{T_0 \cup T_1}\|_2 \bigg(C_1\epsilon\sqrt{NI(1+\delta_{2s})} + \sqrt{2}\delta_{2s} \sum_{j \geq 2} \|\boldsymbol{h}_{T_j}\|_2 \bigg).
\label{4e} \end{align}

\item We have 
\begin{align}
\sum_{j \geq 2}\|\boldsymbol{h}_{T_j}\|_2 & \leq s^{-1/2} \|\boldsymbol{h}_{(T_0)^c}\|_1 \nonumber \\
& \leq s^{-1/2} \|\boldsymbol{h}_{{(T_0)}}\|_1  + 2s^{-1/2}\|\boldsymbol{\theta}-\boldsymbol{\theta_s}\|_1\nonumber \\
& \leq \|\boldsymbol{h}_{{(T_0)}}\|_2 + 2s^{-1/2}\|\boldsymbol{\theta}-\boldsymbol{\theta_s}\|_1  \nonumber\\
& \leq  \|\boldsymbol{h}_{T_0 \cup T_1}\|_2 +  2s^{-1/2}\|\boldsymbol{\theta}-\boldsymbol{\theta_s}\|_1.
\label{4_f1} \end{align}
Combining Eqns. \ref{4e} and \ref{4_f1},
\begin{align}
\|\boldsymbol{h}_{T_0 \cup T_1}\|_2  & \leq  C_1\epsilon \dfrac{\sqrt{NI(1+\delta_{2s}})}{1- (1 + \sqrt{2})\delta_{2s}} + \dfrac{2\sqrt{2}\delta_{2s}}{1- (1 + \sqrt{2})\delta_{2s}} s^{-1/2} \|\boldsymbol{\theta}-\boldsymbol{\theta_s}\|_1.
\label{4f2} \end{align}
\end{enumerate}

\item Combining the upper bounds on $\|\boldsymbol{h}_{(T_0 \cup T_1)}\|_2$ and $\|\boldsymbol{h}_{(T_0 \cup T_1)^c}\|_2$ yields the final result as follows:
\begin{align*} \|\boldsymbol{h}\|_2 &= \|\boldsymbol{h}_{T_0 \cup T_1} + \boldsymbol{h}_{(T_0 \cup T_1)^c}\|_2  \leq \|\boldsymbol{h}_{T_0 \cup T_1}\|_2 + \|\boldsymbol{h}_{(T_0 \cup T_1)^c}\|_2 \leq 2 \|\boldsymbol{h}_{T_0 \cup T_1}\|_2 + 2{s^{-1/2}}\|\boldsymbol{\theta}-\boldsymbol{\theta_s}\|_1.
\end{align*}
Using Eqn. \ref{4f2}, we get
\begin{align*}
\|\boldsymbol{h}\|_2 &\leq  2C_1 \epsilon \dfrac{\sqrt{NI(1+\delta_{2s}})}{1- (1 + \sqrt{2})\delta_{2s}}  + \Big(\dfrac{2 - 2\delta_{2s} + 2 \sqrt{2 \delta_{2s}}}{1- (1 + \sqrt{2})\delta_{2s}}\Big)  s^{-1/2}\|\boldsymbol{\theta}-\boldsymbol{\theta_s}\|_1.
\end{align*}
Let us define $C' \triangleq \dfrac{4\sqrt{8(1+\delta_{2s})}}{\sqrt{p(1-p)}(1- (1 + \sqrt{2})\delta_{2s})}$ and $C'' \triangleq  \Big(\dfrac{2 - 2\delta_{2s} + 2 \sqrt{2 \delta_{2s}}}{1- (1 + \sqrt{2})\delta_{2s}}\Big)$. This yields
\begin{align} \|\boldsymbol{h}\|_2 \leq  C'\sqrt{NI}\epsilon + C'' s^{-1/2} \|\boldsymbol{\theta}-\boldsymbol{\theta_s}\|_1.\end{align}
The positivity requirements for $C'$ and $C''$ are met by $\delta_{2s} < \sqrt{2}-1$. Dividing both sides by $I$ we obtain the first part of the theorem, 
\begin{align*} \dfrac{\|\boldsymbol{\theta - \theta^{\star}}\|_2}{I}  \leq  C'\sqrt{\dfrac{N}{I}}\epsilon + \dfrac{C'' s^{-1/2} \|\boldsymbol{\theta}-\boldsymbol{\theta_s}\|_1}{I}.
\end{align*}
However using tail bounds on $\sqrt{J(\boldsymbol{y},\boldsymbol{\Phi x})}$ from Theorem 1 in Section \ref{subsec:JSD_SQJSD}, we can set $\epsilon = \sqrt{N}(\frac{1}{2}+\frac{\sqrt{11}}{8})$. This yields the following:
\begin{equation} \begin{split}
\textrm{Pr}\Big( \dfrac{\|\boldsymbol{\theta - \theta^{\star}}\|_2}{I}  &\leq  \tilde{C}\dfrac{N}{\sqrt{I}} + \dfrac{C'' s^{-1/2} \|\boldsymbol{\theta}-\boldsymbol{\theta}_s\|_1}{I} \Big) \geq 1-2e^{-N/2},
\end{split} \end{equation}
where $\tilde{C} \triangleq C'(1/2+\sigma)$ where $\sigma$ is the upper bound of $\frac{\sqrt{11}}{8}$ on the standard deviation of the SQJSD as stated in Theorem 1. For high intensity signals, the previous analysis shows that $\sigma$ is independent of both $I$ and $N$. $\Box$
\end{enumerate}

\section{Supplemental Material}
\label{sec:suppmat}
{T}his is supplemental material accompanying the main paper. It basically contains proofs of some lemmas used in the proof of the main theorem in the main paper.

\textbf{Lemma 1} The square root of the Jensen-Shannon Divergence is a metric \cite{Endres2003}.\\
\emph{Proof:} The square root of the Jensen-Shannon divergence trivially obeys the properties of symmetry, non-negativity and identity. We would like to point out that the proof of the triangle inequality for the square-root of the Jensen-Shannon divergence given in \cite{Endres2003} does not require $\boldsymbol{p}$ and $\boldsymbol{q}$ to be probability distributions. In other words given non-negative vectors $\boldsymbol{p}$, $\boldsymbol{q}$, and $\boldsymbol{r}$, we have $\sqrt{J(\boldsymbol{p},\boldsymbol{q})} \leq \sqrt{J(\boldsymbol{p},\boldsymbol{r})} + \sqrt{J(\boldsymbol{q},\boldsymbol{r})}$ even if $\|\boldsymbol{p}\|_1 \neq 1$, $\|\boldsymbol{q}\|_1 \neq 1$ and $\|\boldsymbol{r}\|_1 \neq 1$. We reproduce a sketch of the proof here.
\\
First, we define the function $L(p,q) \triangleq p \log \dfrac{2p}{p+q} + q \log \dfrac{2q}{p+q}$ where scalars $p \in \mathbb{R}_{+}, q \in \mathbb{R}_{+}$. Given any scalar $r \in \mathbb{R}_+$, it is proved in \cite{Endres2003} that $\sqrt{L(p,q)} \leq \sqrt{L(p,r)} + \sqrt{L(q,r)}$. Now, we can clearly see that $\sqrt{J(\boldsymbol{p},\boldsymbol{q})} =  \sqrt{\sum_i L(p_i,q_i)}$. Starting from this, we have
\begin{align*}
\sqrt{J(\boldsymbol{p},\boldsymbol{q})} =  \sqrt{\sum_i L(p_i,q_i)} = \sqrt{\sum_i \bigg( \sqrt{L(p_i,q_i)} \bigg)^2} \leq \sqrt{\sum_i \bigg( \sqrt{L(p_i,r_i)} + \sqrt{L(q_i,r_i)} \bigg)^2} \\
= \sqrt{\sum_i L(p_i,r_i)} + \sqrt{\sum_i L(q_i,r_i)} \textrm{ by Minkowski's inequality } \\
= \sqrt{J(\boldsymbol{p},\boldsymbol{r})} + \sqrt{J(\boldsymbol{q},\boldsymbol{r})}. \Box
\end{align*}

\textbf{Lemma 2:} Let us define
\begin{equation*} \begin{split}
V(\boldsymbol{p},\boldsymbol{q}) \triangleq \sum_{i=1}^{n} |{p_i - q_i}| \\
\Delta(\boldsymbol{p},\boldsymbol{q})  \triangleq \sum_{i=1}^n {\dfrac{{|p_i - q_i|}^2}{p_i + q_i}}. \nonumber
\end{split} \end{equation*}
If $\boldsymbol{p}, \boldsymbol{q} \succeq 0$ and $\|\boldsymbol{p}\|_1 \leq 1$, $\|\boldsymbol{q}\|_1 \leq 1$ then
\begin{equation}  \dfrac{1}{2}V(\boldsymbol{p},\boldsymbol{q})^2 \leq \Delta(\boldsymbol{p},\boldsymbol{q})  \leq 4 J(\boldsymbol{p},\boldsymbol{q}).\end{equation}\\
\emph{Proof:} The latter inequality can be proved using arguments in \cite{Topsoe2000} (Section III) as these arguments do not require $\boldsymbol{p}$ and $\boldsymbol{q}$ to be probability distributions in any of the steps. To prove the first inequality, we prove that $2\Delta (\boldsymbol{p},\boldsymbol{q}) - V(\boldsymbol{p},\boldsymbol{q})^2  \geq 0$ as follows. Let us define $z_i \triangleq |p_i - q_i|$ and $w_i \triangleq \dfrac{1}{2}|p_i + q_i|$. If $\|\boldsymbol{p}\|_1 \leq 1$, $\|\boldsymbol{q}\|_1 \leq 1$ then $\sum_{i=1}^n w_i \leq 1$. Hence $\exists \alpha \geq  0$ such that $\sum_{i=1}^n w_i + \alpha = 1$.
\begin{equation} \begin{split}
2\Delta(\boldsymbol{p},\boldsymbol{q}) -V(\boldsymbol{p},\boldsymbol{q})^2 \\
= \sum_{i=1}^n \dfrac{{z_i}^2}{w_i} - \big(\sum_{i=1}^n z_i \big)^2 \\
= \dfrac{1}{\prod_{i=1}^n w_i} \Bigg[\sum_{i=1}^n {z_i}^2 \prod_{j \neq i} w_j - ({\prod_{i=1}^n w_i}) (\sum_{i=1}^n z_i)^2 \Bigg] \\
= \dfrac{1}{\gamma} \Bigg[\sum_{i=1}^n {z_i}^2 \prod_{j \neq i} w_j - \gamma \big(\sum_{i=1}^n z_i \big)^2 \Bigg] \text{ where } \gamma \triangleq {\prod_{i=1}^n w_i}\\
= \dfrac{1}{\gamma}\Bigg[\sum_{i=1}^n {z_i}^2 \prod_{j \neq i}w_j \big(1-w_i\big) - 2\big(\sum_{i} \sum_{j > i} z_i z_j \big) \gamma \Bigg] \\
= \dfrac{1}{\gamma}\Bigg[\sum_{i=1}^n {z_i}^2 \prod_{j \neq i}w_j \big(\sum_{j\neq i}w_j + \alpha \big) - 2\big(\sum_{i} \sum_{j > i} z_i z_j \big) \gamma \Bigg] \\
= \dfrac{1}{\gamma}\Bigg[(\sum_{i} \sum_{j > i}\big(z_i w_j - z_j w_i \big)^2\prod_{k \neq j, k \neq i}w_k) + \big(\alpha \sum_{i=1}^n {z_i}^2 \prod_{j \neq i} w_j \big)\Bigg] \geq 0
\end{split} \end{equation}
Notice that the first term in the last step is clearly non-negative as it is the product of a square-term and a term containing $w_k$ values all of which are non-negative and since $\gamma \geq 0$. The second term is also non-negative as $\alpha \geq 0$. Thus, the inequality $ \dfrac{1}{2}V(\boldsymbol{p},\boldsymbol{q})^2 \leq \Delta (\boldsymbol{p},\boldsymbol{q})$ is proved. $\Box$

\textbf{Lemma 3:} Given non-negative vectors $\boldsymbol{u}$ and $\boldsymbol{v}$, we have $\dfrac{1}{4}D_s(\boldsymbol{u},\boldsymbol{v}) \geq J(\boldsymbol{u},\boldsymbol{v})$.\\
\textit{Proof:} Following \cite{Lin1991}, we have $\dfrac{u_i + v_i}{2} \geq \sqrt{u_i v_i}$ by the arithmetic-geometric inequality. Now we have:
\begin{eqnarray}
J(\boldsymbol{u},\boldsymbol{v}) = \dfrac{1}{2}(\sum_{i} u_i \log \dfrac{u_i}{(u_i+v_i)/2} + v_i \log \dfrac{v_i}{(u_i+v_i)/2}) \\ \nonumber
\leq \dfrac{1}{2}( \sum_{i} u_i \log \dfrac{u_i}{\sqrt{u_i v_i}} + v_i \log \dfrac{v_i}{\sqrt{u_i v_i}}) \leq \dfrac{1}{4}(\sum_{i} u_i \log \dfrac{u_i}{v_i} + v_i \log \dfrac{v_i}{u_i}) = \dfrac{1}{4}D_s(\boldsymbol{u},\boldsymbol{v}). \\
\label{eq:D_s_J}
\end{eqnarray}
In \cite{Lin1991}, this proof is presented for probability mass functions, but we observe here that it extends to arbitrary non-negative vectors. $\Box$

\textbf{Lemma 4:} Given $\boldsymbol{\theta}$ which is the minimizer of problem (\textsf{P4}) for some $\lambda > 0$, there exists some value of $\epsilon = \epsilon_\theta$ for which $\boldsymbol{\theta}$ is the minimizer of problem (\textsf{P2}), but without the constraint $\|\boldsymbol{\Psi \theta}\|_1 = I$.
\emph{Proof:}
Our proof follows \cite{Foucart2013}, proposition 3.2. Define $\epsilon_\theta \triangleq J(\boldsymbol{\Phi \Psi \theta},\boldsymbol{y})$. Consider vector $\boldsymbol{\theta'}$ such that $J(\boldsymbol{\Phi \Psi \theta'},\boldsymbol{y}) \leq \epsilon_\theta$. Now since $\boldsymbol{\theta}$ minimizes (\textsf{P3}), we have $\lambda \|\boldsymbol{\theta}\|_1 + J(\boldsymbol{\Phi \Psi \theta},\boldsymbol{y}) \leq \lambda \|\boldsymbol{\theta'}\|_1 + J(\boldsymbol{\Phi \Psi \theta'},\boldsymbol{y}) \leq \lambda \|\boldsymbol{\theta'}\|_1 + J(\boldsymbol{\Phi \Psi \theta},\boldsymbol{y})$, yielding $\|\boldsymbol{\theta}\|_1 \leq \|\boldsymbol{\theta'}\|_1$, thereby establishing that $\boldsymbol{\theta}$ is also the minimizer of a version of (\textsf{P2}) without the constraint $\|\boldsymbol{\Psi \theta}\|_1 = I$. $\Box$

\section*{References}
\bibliography{refs}

\end{document}

\begin{frontmatter}

\title{Supplemental Material: Reconstruction Error Bounds for Compressed Sensing under Poisson Noise using the Square Root of the Jensen-Shannon Divergence}

%\author{Sukanya Patil and Ajit Rajwade\fnref{myfootnote}}

\author[mymainaddress]{Sukanya Patil}
\ead{sukanyapatil1993@gmail.com}
\address[mymainaddress]{Department of Electrical Engineering, IIT Bombay}

\author[mymainaddress2]{Ajit Rajwade\corref{mycorrespondingauthor}}
\cortext[mycorrespondingauthor]{Corresponding author}
\ead{ajitvr@cse.iitb.ac.in}
\address[mymainaddress2]{Department of Computer Science and Engineering, IIT Bombay}
\fntext[myfootnote]{AR gratefully acknowledges support from IIT Bombay Seed Grant number 14IRCCSG012.}

\end{frontmatter}

\linenumbers

{T}his is supplemental material accompanying the main paper. It basically contains proofs of some lemmas used in the proof of the main theorem in the main paper.

\textbf{Lemma 1} The square root of the Jensen-Shannon Divergence is a metric \cite{Endres2003}.\\
\emph{Proof:} The square root of the Jensen-Shannon divergence trivially obeys the properties of symmetry, non-negativity and identity. We would like to point out that the proof of the triangle inequality for the square-root of the Jensen-Shannon divergence given in \cite{Endres2003} does not require $\boldsymbol{P}$ and $\boldsymbol{Q}$ to be probability distributions. In other words given non-negative vectors $\boldsymbol{P}$, $\boldsymbol{Q}$, and $\boldsymbol{R}$, we have $\sqrt{J(\boldsymbol{P},\boldsymbol{Q})} \leq \sqrt{J(\boldsymbol{P},\boldsymbol{R})} + \sqrt{J(\boldsymbol{Q},\boldsymbol{R})}$ even if $\|\boldsymbol{P}\|_1 \neq 1$, $\|\boldsymbol{Q}\|_1 \neq 1$ and $\|\boldsymbol{R}\|_1 \neq 1$. We reproduce a sketch of the proof here.
\\
First, we define the function $L(p,q) \triangleq p \log \dfrac{2p}{p+q} + q \log \dfrac{2q}{p+q}$ where scalars $p \in \mathbb{R}_{+}, q \in \mathbb{R}_{+}$. Given any scalar $r \in \mathbb{R}_+$, it is proved in \cite{Endres2003} that $\sqrt{L(p,q)} \leq \sqrt{L(p,r)} + \sqrt{L(q,r)}$. Now, we can clearly see that $\sqrt{J(\boldsymbol{P},\boldsymbol{Q})} =  \sqrt{\sum_i L(p_i,q_i)}$. Starting from this, we have
\begin{align*}
\sqrt{J(\boldsymbol{P},\boldsymbol{Q})} =  \sqrt{\sum_i L(p_i,q_i)} \\
= \sqrt{\sum_i \bigg( \sqrt{L(p_i,q_i)} \bigg)^2} \\
\leq \sqrt{\sum_i \bigg( \sqrt{L(p_i,r_i)} + \sqrt{L(q_i,r_i)} \bigg)^2} \\
= \sqrt{\sum_i L(p_i,r_i)} + \sqrt{\sum_i L(q_i,r_i)} \textrm{ by Minkowski's inequality } \\
= \sqrt{J(\boldsymbol{P},\boldsymbol{R})} + \sqrt{J(\boldsymbol{Q},\boldsymbol{R})}. \Box
\end{align*}

\textbf{Lemma 2:} Let us define
\begin{equation*} \begin{split}
V(\boldsymbol{P},\boldsymbol{Q}) \triangleq \sum_{i=1}^{n} |{p_i - q_i}| \\
\Delta(\boldsymbol{P},\boldsymbol{Q})  \triangleq \sum_{i=1}^n {\dfrac{{|p_i - q_i|}^2}{p_i + q_i}}. \nonumber
\end{split} \end{equation*}
If $\boldsymbol{P}, \boldsymbol{Q} \succeq 0$ and $\|\boldsymbol{P}\|_1 \leq 1$, $\|\boldsymbol{Q}\|_1 \leq 1$ then
\begin{equation}  \dfrac{1}{2}V(\boldsymbol{P},\boldsymbol{Q})^2 \leq \Delta(\boldsymbol{P},\boldsymbol{Q})  \leq 4 J(\boldsymbol{P},\boldsymbol{Q}).\end{equation}\\
\emph{Proof:} The latter inequality can be proved using arguments in \cite{Topsoe2000} (Section III) as these arguments do not require $\boldsymbol{P}$ and $\boldsymbol{Q}$ to be probability distributions in any of the steps. To prove the first inequality, we prove that $2\Delta (\boldsymbol{P},\boldsymbol{Q}) - V(\boldsymbol{P},\boldsymbol{Q})^2  \geq 0$ as follows. Let us define $z_i \triangleq |p_i - q_i|$ and $w_i \triangleq \dfrac{1}{2}|p_i + q_i|$. If $\|\boldsymbol{P}\|_1 \leq 1$, $\|\boldsymbol{Q}\|_1 \leq 1$ then $\sum_{i=1}^n w_i \leq 1$. Hence $\exists \alpha \geq  0$ such that $\sum_{i=1}^n w_i + \alpha = 1$.
\begin{equation} \begin{split}
2\Delta(\boldsymbol{P},\boldsymbol{Q}) -V(\boldsymbol{P},\boldsymbol{Q})^2 \\
= \sum_{i=1}^n \dfrac{{z_i}^2}{w_i} - \big(\sum_{i=1}^n z_i \big)^2 \\
= \dfrac{1}{\prod_{i=1}^n w_i} \Bigg[\sum_{i=1}^n {z_i}^2 \prod_{j \neq i} w_j - ({\prod_{i=1}^n w_i}) (\sum_{i=1}^n z_i)^2 \Bigg] \\
= \dfrac{1}{\gamma} \Bigg[\sum_{i=1}^n {z_i}^2 \prod_{j \neq i} w_j - \gamma \big(\sum_{i=1}^n z_i \big)^2 \Bigg] \text{ where } \gamma \triangleq {\prod_{i=1}^n w_i}\\
= \dfrac{1}{\gamma}\Bigg[\sum_{i=1}^n {z_i}^2 \prod_{j \neq i}w_j \big(1-w_i\big) - 2\big(\sum_{i} \sum_{j > i} z_i z_j \big) \gamma \Bigg] \\
= \dfrac{1}{\gamma}\Bigg[\sum_{i=1}^n {z_i}^2 \prod_{j \neq i}w_j \big(\sum_{j\neq i}w_j + \alpha \big) - 2\big(\sum_{i} \sum_{j > i} z_i z_j \big) \gamma \Bigg] \\
= \dfrac{1}{\gamma}\Bigg[(\sum_{i} \sum_{j > i}\big(z_i w_j - z_j w_i \big)^2\prod_{k \neq j, k \neq i}w_k) + \big(\alpha \sum_{i=1}^n {z_i}^2 \prod_{j \neq i} w_j \big)\Bigg]\\
\geq 0
\end{split} \end{equation}
Notice that the first term in the last step is clearly non-negative as it is the product of a square-term and a term containing $w_k$ values all of which are non-negative and since $\gamma \geq 0$. The second term is also non-negative as $\alpha \geq 0$. Thus, the inequality $ \dfrac{1}{2}V(\boldsymbol{P},\boldsymbol{Q})^2 \leq \Delta (\boldsymbol{P},\boldsymbol{Q})$ is proved. $\Box$

\textbf{Lemma 3:} Given non-negative vectors $\boldsymbol{u}$ and $\boldsymbol{v}$, we have $\dfrac{1}{4}D_s(\boldsymbol{u},\boldsymbol{v}) \geq J(\boldsymbol{u},\boldsymbol{v})$.\\
\textit{Proof:} Following \cite{Lin1991}, we have $\dfrac{u_i + v_i}{2} \geq \sqrt{u_i v_i}$ by the arithmetic-geometric inequality. Now we have:
\begin{eqnarray}
J(\boldsymbol{u},\boldsymbol{v}) = \dfrac{1}{2}(\sum_{i} u_i \log \dfrac{u_i}{(u_i+v_i)/2} + v_i \log \dfrac{v_i}{(u_i+v_i)/2}) \\ \nonumber
\leq \dfrac{1}{2}( \sum_{i} u_i \log \dfrac{u_i}{\sqrt{u_i v_i}} + v_i \log \dfrac{v_i}{\sqrt{u_i v_i}}) \\ \nonumber
\leq \dfrac{1}{4}(\sum_{i} u_i \log \dfrac{u_i}{v_i} + v_i \log \dfrac{v_i}{u_i}) \\ \nonumber
= \dfrac{1}{4}D_s(\boldsymbol{u},\boldsymbol{v}). \\
\label{eq:D_s_J}
\end{eqnarray}
In \cite{Lin1991}, this proof is presented for probability mass functions, but we observe here that it extends to arbitrary non-negative vectors. $\Box$

\\
\textbf{Lemma 4:} Given $\boldsymbol{\theta}$ which is the minimizer of problem (\textsf{P4}) for some $\lambda > 0$, there exists some value of $\epsilon = \epsilon_\theta$ for which $\boldsymbol{\theta}$ is the minimizer of problem (\textsf{P2}), but without the constraint $\|\boldsymbol{\Psi \theta}\|_1 = I$.\\
\emph{Proof:}
Our proof follows \cite{Foucart2013}, proposition 3.2. Define $\epsilon_\theta \triangleq J(\boldsymbol{\Phi \Psi \theta},\boldsymbol{y})$. Consider vector $\boldsymbol{\theta'}$ such that $J(\boldsymbol{\Phi \Psi \theta'},\boldsymbol{y}) \leq \epsilon_\theta$. Now since $\boldsymbol{\theta}$ minimizes (\textsf{P3}), we have $\lambda \|\boldsymbol{\theta}\|_1 + J(\boldsymbol{\Phi \Psi \theta},\boldsymbol{y}) \leq \lambda \|\boldsymbol{\theta'}\|_1 + J(\boldsymbol{\Phi \Psi \theta'},\boldsymbol{y}) \leq \lambda \|\boldsymbol{\theta'}\|_1 + J(\boldsymbol{\Phi \Psi \theta},\boldsymbol{y})$, yielding $\|\boldsymbol{\theta}\|_1 \leq \|\boldsymbol{\theta'}\|_1$, thereby establishing that $\boldsymbol{\theta}$ is also the minimizer of a version of (\textsf{P2}) without the constraint $\|\boldsymbol{\Psi \theta}\|_1 = I$. $\Box$

\section*{References}
\bibliography{refs}